\documentclass[twocolumn,pre,superscriptaddress,bibnotes,notitlepage,nofootinbib]{revtex4-2}

\usepackage{CJK}

\usepackage{graphicx}

\usepackage{dcolumn}

\usepackage{bm}
\usepackage[normalem]{ulem}
\usepackage{color}
\usepackage{amsmath}
\usepackage{amssymb}
\usepackage{lipsum}
\usepackage{bm}
\usepackage{xcolor}
\usepackage{graphicx}
\usepackage{verbatim}
\usepackage[T1]{fontenc}
\usepackage{hyperref}
\usepackage{comment}
\usepackage{mathtools}
\makeatletter
\DeclareRobustCommand{\smvdots}{%
  \vbox{%
    \baselineskip=2.5pt 
    \lineskiplimit=0pt
    \kern3pt
    \hbox{.}\hbox{.}\hbox{.}%
  }%
}
\makeatother
\numberwithin{equation}{subsection}
\renewcommand{\theequation}{\thesection.\thesubsection.\arabic{equation}}

\ExplSyntaxOn
\NewDocumentEnvironment{multiequation}{b}
 {
  \vantiempham:n { #1 }
 }
 {}

\seq_new:N \l__vantiempham_md_seq

\cs_new_protected:Nn \vantiempham:n
 {
  \seq_set_split:Nnn \l__vantiempham_md_seq { \\ } { #1 }
  $$
  \seq_map_function:NN \l__vantiempham_md_seq \__vantiempham_md_item:n
  $$
 }
\cs_new_protected:Nn \__vantiempham_md_item:n
 {
  \begin{minipage}{\dim_eval:n {\displaywidth/(\seq_count:N \l__vantiempham_md_seq)} }
  \begin{equation}
  \cs_set_eq:Nc \label { ltx@label }
  #1
  \vphantom{\seq_use:Nn \l__vantiempham_md_seq {}}
  \end{equation}
  \end{minipage}
 }

\ExplSyntaxOff  

\newcommand{\bsf}[1]{\textsf{\textbf{#1}}}
\usepackage{epstopdf}

\newcommand{\beq}{\begin{equation}}
	\newcommand{\eeq}{\end{equation}}
\newcommand{\beqn}{\begin{eqnarray}}
	\newcommand{\eeqn}{\end{eqnarray}}

\newcommand{\bew}{\begin{widetext}}
	\newcommand{\ew}{\end{widetext}}

\usepackage{xcolor}
\definecolor{CMaroon}{HTML}{800000}

\begin{document}
	\title{Effect of chirality on active elastomers and solids in three dimensions} 
    \author{S. J. Kole}
\email{sjkole.acad@gmail.com}
\affiliation{Pritzker School of Molecular Engineering, University of Chicago, Chicago, Illinois 60637, USA}

\author{Gareth P. Alexander}
\email{G.P.Alexander@warwick.ac.uk}
\affiliation{Department of Physics, Gibbet Hill Road, University of Warwick, Coventry CV4 7AL, United Kingdom}

\author{Ananyo Maitra}
\email{nyomaitra07@gmail.com}
\affiliation{{Laboratoire de Physique Th\'eorique et Mod\'elisation, CNRS UMR 8089,
		CY Cergy Paris Universit\'e, F-95032 Cergy-Pontoise Cedex, France}}
\affiliation{Sorbonne Universit\'{e} and CNRS, Laboratoire Jean Perrin, F-75005, Paris, France}

\author{Sriram Ramaswamy}
\email{sriram@iisc.ac.in}
\affiliation{Centre for Condensed Matter Theory, Department of Physics, Indian Institute of Science, Bangalore 560 012, India}
\affiliation{International Centre for Theoretical Sciences, Tata Institute of Fundamental Research, Bangalore 560 089, India}

	
\begin{abstract}
We examine the effect of chiral activity on nematic and polar uniaxial elastomers and solids by constructing the general hydrodynamic equations for these phases. We start with the elastomer, which is not only invariant under rigid rotations but also under an independent rotation of the uniaxial direction, and show that this higher symmetry strongly constrains the effects of chiral activity. We then break the elastomeric symmetry by adding a symmetry-breaking term to the free energy but still retain invariance under rigid rotations. The resulting state has the symmetry of a uniaxial solid, and we show that in a fixed background medium chiral activity generically destabilises it. The force density responsible for this {{instability}} vanishes as the field that breaks the elastomeric symmetry goes to zero. Further, the presence of the symmetry-breaking field also allows for an additional odd elastic force density that generically destabilises momentum-conserved chiral and uniaxial solids in the inertial regime.
\end{abstract}
	
\maketitle
\section{Introduction}

Chirality is the basic property of materials or geometrical figures corresponding to their two mirror images being distinguishable~\cite{Kelvin}. It is fundamentally dimension-dependent: objects that are chiral in two dimensions are not chiral in three because they can be flipped over and converted to their mirror image through the third dimension. Therefore, it is natural that the physics of chiral materials should also be dimension-dependent. This is already known in systems at thermal equilibrium: while objects that are chiral in two dimensions---such as an `L'-shaped particle---do not have any special phase behaviour, objects that are chiral in three dimensions form a wide range of spectacular chiral phases with twisted, periodic morphologies~\cite{deGen,Mermin_blue,pisljar2022,renn1988,hough2009chiral,hough2009helical,matsumoto2017straight} and support an enormous variety of topological textures in chiral liquid crystals~\cite{chen2013,machon2017,pollard2019,pollard2023,tai2019,wu2022hopfions} and magnets~\cite{roessler2006spontaneous,sutcliffe2018hopfions,kent2021creation,chen2026laser,li2026electrically}. While these structures \emph{form} due to chirality, the effect of parity asymmetry is not apparent in the leading-order hydrodynamic mode structure {and arises only as} next-to-leading-order corrections to hydrodynamics~\cite{Alexander_chol}.

In active systems~\cite{RMP, ramaswamy2010mechanics}, the situation is very different. Two-dimensional chirality allows for new phases not possible in passive materials, dramatically modifies the hydrodynamic mode structure in ordered phases and can generically destabilise a symmetry-broken state. The time cholesteric state~\cite{maitra2019spontaneous, AM_hex, AM_chi_rev, Toner_rot} is an example of the first, sound modes in inertialess solids due to odd elasticity~\cite{scheibner2020odd, fruchart2023odd, AM_hex, AM_chi_rev} is an example of the second, and the generic instability of a layered phase in momentum-conserved systems due to chiral activity~\cite{SJ_smec, AM_chi_rev} is an example of the third. In two dimensions, chiral active effects arise already in isotropic systems~\cite{scheibner2020odd}, or hexagonal crystals~\cite{bililign2022motile,veenstra2025adaptive, kole2025non}, whose elasticity is isotropic, and in spinning active Brownian particles~\cite{liebchen2022chiral}. These provide an idealised minimal setting for exploring chiral active phenomena at continuum and particle levels, respectively. A corresponding minimal model in three dimensions is not so immediate: isotropy precludes any odd elasticity in three-dimensional isotropic solids~\cite{scheibner2020odd}; the displacement field dynamics of active smectics is unaffected by chirality to leading order in gradients~\cite{SJ_smec, AM_chi_rev}; and chiral polar columnar materials have an odd elastic force density that has the same form as the one in two-dimensional solids~\cite{SJ_col, AM_chi_rev}. In this article and in the accompanying Letter~\cite{short}, we examine the effect of chiral activity on uniaxial–––both polar and nematic–––three-dimensional solids and elastomers and show that their dynamics is as significant, and arguably richer, than their two-dimensional counterparts.  

The properties of three-dimensional active chiral solids should be of relevance to a wide range of biological systems, such as living tissues moving through an extracellular matrix~\cite{rustarazo2026adhesion,spatarelu2019biomechanics,rorth2009collective,ellison2016cell,mugler2016limits,camley2016emergent} and twisted fibrous structures~\cite{Bouligand_LC,Bouligand_twisted,Heartstrings}. Indeed, the form of the latter class of materials resembles the likely end state of the instability of chiral active polar solids that we discuss {below}. The study of three-dimensional active chiral solids is so far relatively neglected and yet of both fundamental and practical importance given that life is three-dimensional and uncompromisingly chiral.

In the accompanying Letter~\cite{short}, we have shown, based on general symmetry arguments, that chiral and active force densities generically destabilise three-dimensional uniaxial solids in a frictional medium. This instability is caused by a chiral active force that is lower order in gradients than odd elasticity and equally arises in an isotropic chiral solid moving in an anisotropic medium. We also argued that a higher-symmetry phase---a uniaxial elastomer---remains stable in chiral active systems. This is because its higher symmetry eliminates the force densities responsible for instability in the solid.

In this long paper, we construct the equations of motion for the dynamics of chiral and active polar (and nematic) elastomers, within a formalism that describes the breaking of orientational symmetry in isotropic elastomers. This allows a systematic and unified treatment of chiral, active elastomers and solids in three dimensions that explicitly incorporates their symmetries. We show that the elastomeric symmetry under independent rotation of the orientation direction and the solid structure eliminates certain contributions to the equations of motion present in more general uniaxial solids. We then explicitly show that breaking the additional symmetry---which is a natural consequence of crosslinking---restores the missing contributions, thereby eliminating the phase due to the chiral active instability. We also discuss the effect of breaking additional symmetries, for instance, due to an external field or a structured medium. General representation-theory arguments show that polar uniaxial solids have three odd elasticities. We show how these arise within our formalism, including a non-chiral and apolar odd elasticity~\cite{zahalak1996non,shankar2024active}, and show that one of the chiral odd elasticities is eliminated by the higher symmetry of elastomers. Finally, we display the equations of motion for the displacement fields for chiral and active uniaxial elastomers and solids in various dynamical regimes and discuss their linear fluctuation spectra. 

This paper is organised as follows: in Sec.~\ref{sec:Actelforce}, we construct passive and active elastic force densities for elastomers in terms of a strain defined with respect to an isotropic reference state and orientational order parameter. The terms discussed and collected in this section appear in what amounts to an ``order parameter'' theory of chiral active elastomers. Orientationally ordered elastomers appear as symmetry-broken states of this theory. Our goal in later sections is to study the Nambu-Goldstone mode dynamics in these lower-symmetry ordered states.
In Sec.~\ref{sec:vforceden}, we discuss velocity-dependent force densities that can arise in polar and uniaxial systems. In Sec.~\ref{sec:passens}, we primarily recapitulate rotation-symmetry breaking in passive elastomers and the physics of the symmetry-broken state, as a prelude to discussing the effect of activity on such phases. We also discuss why uniaxial states can be linearly stable chiral solids while they are generically linearly unstable in chiral fluids. In Sec.~\ref{sec:chielexp}, we discuss the effect of chiral activity on polar elastomers. We show that the symmetry of elastomers under independent rotations of the elastic material and the orientation direction eliminate certain terms that arise in theories of solids, which are invariant only under the joint rotation of the two. We then discuss odd elasticities in this system before constructing the dynamical equations for the displacement fields---the only Nambu-Goldstone modes for polar and nematic elastomers---in a variety of physical settings: in a fixed background medium, in an incompressible, Stokesian fluid medium and in the absence of any dissipation. In Sec.~\ref{sec:Unisol}, we discuss the effect of chiral activity on polar solids. We arrive at the dynamics of polar, chiral and active solids by starting with the theory of elastomers and explicitly breaking the higher symmetry by introducing a field. We then construct the dynamical equations for the displacement fields in the same physical regimes discussed in Sec. ~\ref{sec:chielexp}. Finally, we present a general discussion of elasticity of uniaxial solids using group representations. In Sec.~\ref{sec:extenviron}, we discuss the dynamics of solids and elastomers in a fixed background medium where the medium is not isotropic, but itself provides a set of preferred directions for the solid. We close in Sec.~\ref{sec:conc} by summarising our results.

\section{Active elastic forces and stresses}
\label{sec:Actelforce}
In this section, we construct the active elastic force densities and stresses for elastomers in terms of the strain tensor with respect to an isotropic reference state and polarisation. That is, we construct a theory of the elastomeric solid in terms of the polar (or nematic) order parameter that is explicitly invariant under rotations in the \emph{isotropic} state. We will then use this description to construct the dynamics of the uniaxial elastomeric state in which rotation symmetry is spontaneously broken, and the corresponding order parameters have non-zero average amplitudes. We could have used our knowledge of the invariance properties of the nematic or polar elastomeric state---it must remain invariant under \emph{independent} rotations of the solid structure and the ordering direction in contrast to uniaxial solids which remain invariant only under \emph{joint} rotation of both~\cite{Olmsted}---to directly construct the soft Nambu-Goldstone mode (or phononic) dynamics of this state, as one can do for flocks~\cite{AM_floc}. Indeed, we constructed the dynamics of polar and chiral active solids in this manner in~\cite{short}. However, given the complicated nature of this invariance, it is clearer and technically simpler to adopt the strategy that we do here. 

The rest of this section is organised as follows: in Sec.~\ref{subsec:var}, we define the variables in terms of which we construct our forces and stresses. In Sec.~\ref{subsec:active_elastic_stresses} we list the active force densities that we will consider that can be written as divergences of a symmetric tensor and, therefore, are allowed in systems that conserve momentum and angular momentum. In Sec.~\ref{frcactfric} we list the force densities that cannot be so written but affect the long-time, large-scale dynamics of systems we consider when they are in a fixed background medium. 

\subsection{Definition of variables}
\label{subsec:var}
\begin{figure}[t]
    \centering
    \includegraphics[width=\linewidth]{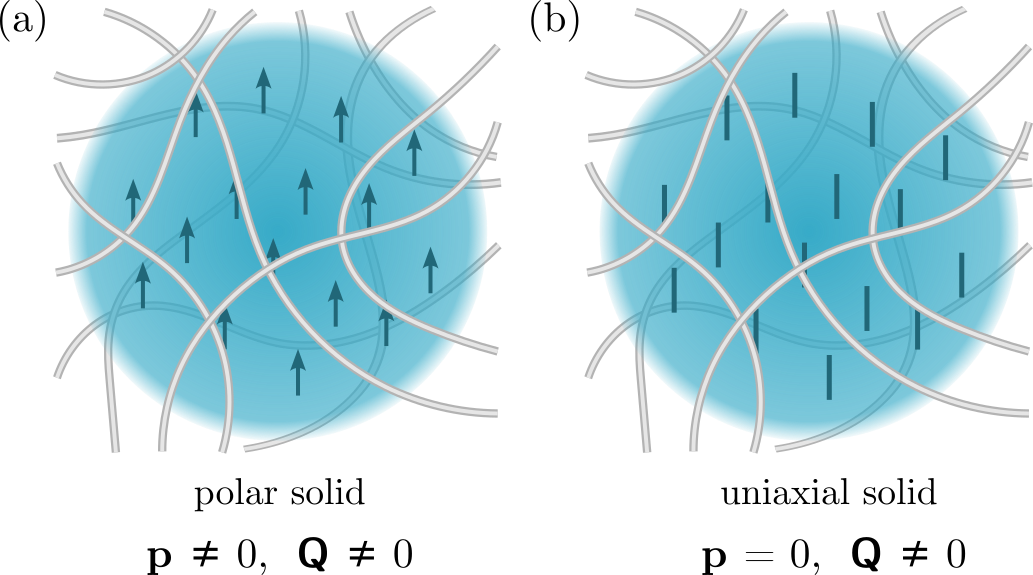}
    \caption{Schematic illustration of oriented solids (blue) moving through an isotropic external medium, visualised as a light grey mesh. (a) Polar solid with both ${\bf p} \ne 0$ and ${\bsf Q} \ne 0$. (b) Uniaxial solid with ${\bf p} = 0$ but ${\bsf Q} \ne 0$.} 
    \label{fig:solids_schematic}
\end{figure}

The polar order parameter that accounts for the degree of polarisation is ${\bf p}({\bf r}, t)$, while the nematic order parameter measuring the degree of uniaxiality is ${\bsf Q}({\bf r},t)$. ${\bsf Q}\neq 0$ and ${\bf p}={\bf{0}}$ in a nematic state but ${\bf p}\neq \bf{0}$ and ${\bsf Q}={\bf pp}-(1/3)p^2{\bsf I}$, where ${\bsf I}$ is the identity tensor and $p=|{\bf p}|\neq 0$, in a polar state, see Fig.~\ref{fig:solids_schematic}. We will primarily consider polar materials in this article; however, whenever relevant, we indicate the terms that are invariant under the inversion of the polar direction ${\bf p}\to-{\bf p}$ and thus also allowed in systems with nematic asymmetry. We also define a unit magnitude polar director $\hat{{\bf n}}$ via ${\bf p}=p\hat{{\bf n}}$. Further, since we are interested in materials composed of three-dimensional chiral elements, stresses and force densities can depend on the rank-3 Levi-Civita tensor $\epsilon_{ijk}$. 

Since we are interested in the linearised displacement dynamics in this article, we define a linearised strain field ${\bsf U}={\bsf C}+(1/3)\Phi{\bsf I}$, $U_{ij}=(\partial_j\bar{u}_i+\partial_i\bar{u}_j)/2$, with $\bar{{\bf u}}$ being a displacement field about an \emph{isotropic} reference state and ${\bsf C}$ and $\Phi$ being the trace-removed and isotropic parts of ${\bsf U}$. At this level of description, there is no distinction between an Eulerian or Lagrangian strain tensor. A nonlinear theory of uniaxial solids and elastomers with chiral and active force densities can be constructed by replacing ${\bsf U}$ with the left Cauchy-Green tensor \cite{Ranjan_Lubensky} $\bar{{\bsf U}}=(1/2)(\partial_j\bar{u}_i+\partial_i\bar{u}_j+\partial_k\bar{u}_i\partial_k\bar{u}_j)$, within a Lagrangian description. 

We consider permanently crosslinked materials such that there is no relative motion between mass and structure. Therefore, the fluctuations $\delta \rho$ of the density field $\rho$ are slaved to {structural dilations, $\delta\rho/\rho_0=-\nabla\cdot\bar{{\bf u}}$,} and play no further role in our analysis. A further consequence of being permanently crosslinked is that the evolution equation for the displacement field is $\dot{\bar{{\bf u}}}={\bf v}$, where the overdot represents the material derivative and ${\bf v}$ is the velocity field, with kinematic contributions proportional to $\nabla \times {\bf v}$ at next order in gradients when the structure is chiral~\cite{Alexander_chol}. Thus, to complete our dynamical description, we only need to construct the equation for force densities.

\subsection{Active force densities that can be expressed as divergences of stresses}
\label{subsec:active_elastic_stresses}
In systems that conserve momentum, all force densities must be expressible as a divergence of a second-rank stress tensor. We will discuss passive contributions to the stress tensor in the next section; here, we construct the active part of the stress tensor $\bm{\sigma}^a$ that depends on the strain and ${\bf p}$. This can be further separated into a chiral $\bm{\sigma}^{ac}$ and an achiral $\bm{\sigma}^{anc}$ part, such that $\bm{\sigma}^a=\bm{\sigma}^{ac}+\bm{\sigma}^{anc}$. For the latter, we retain only one term: 
\begin{equation}
\label{actacstrs}
\sigma_{ij}^{anc}=\zeta_1 Q_{ij}\,.
\end{equation}
We do not consider possible additional active stresses $\propto$ $C_{ij}$, $\Phi\delta_{ij}$, $Q_{ik}C_{kj}+C_{ik}Q_{kj}$ etc. \emph{All} of these stresses (and infinitely many others) yield force densities at $\mathcal{O}(\nabla^2)$---i.e., at the same order in gradients as the term retained in \eqref{actacstrs}---in the nematic elastomeric phase. Since we are interested in a phase in which $\langle {\bsf Q}\rangle\neq 0$, an expansion in powers of ${\bsf Q}$ is not justifiable. However, Eq.~\eqref{actacstrs}, in conjunction with passive stresses discussed later, yields the most generic form of the linearised achiral stress in uniaxial elastomeric phases, and the effect of any additional contributions can be absorbed in redefinitions of phenomenological coefficients. For instance, the contributions $\propto$ $C_{ij}$, $\Phi\delta_{ij}$ simply renormalise the isotropic bulk and shear moduli. However, the fact that we could have retained additional contributions in~\eqref{actacstrs} without modifying the form of the linearised dynamics of the uniaxial elastomeric phase, implies that the relation between the coefficients in~\eqref{actacstrs} and those that appear in dynamical equations for the elastomeric state are \emph{not} universal. Instead, the latter should be seen as independent phenomenological coefficients whose values can be obtained from measurements. 

Note further that~\eqref{actacstrs} is invariant under ${\bf p}\to -{\bf p}$. Like the active stress in fluids, active terms are insensitive to polarity, to leading order in gradients, even in \emph{achiral} elastomers.

For the chiral and active stress, we retain two contributions: 
\begin{equation}
\sigma^{ac}_{ij}=2\left\{\epsilon_{ilk}p_k\left[2\zeta^c_1p_jp_mC_{lm}+\zeta^c_2\left(p_mp_mC_{lj}-p_jp_mC_{lm}\right)\right]\right\}^S \,,
\label{cstrs}
\end{equation}
where the superscript $^S$ denotes the symmetric part of the tensor. The forms of these terms are inspired by our discussion of odd viscosity in three-dimensional chiral fluids~\cite{SJ_col}. We show below that they produce force densities at ${\cal O}(\nabla^2)$ that correspond to the two chiral odd elasticities of polar uniaxial solids. Note that unlike the achiral stress, these are \emph{not} invariant under ${\bf p}\to -{\bf p}$; that is, these stresses are allowed only in polar systems and not in nematic ones. As discussed in~\cite{SJ_smec, SJ_col} uniaxial but apolar contributions to the chiral active stress arise only at the next order in gradients.

\subsection{Active force densities that cannot be expressed as divergence of stresses}
\label{frcactfric}

When a solid or elastomer does not conserve momentum but moves through a fixed background medium, which we assume to be isotropic here, the active force densities need not arise as the divergence of a stress~\cite{Fernando_mem, Ano_pol, Ano_apol}. It is well known that in achiral but polar solids~\cite{Rangan, AM_sol, AM_Ladoux_PNAS}, such forces dominate over passive and active force densities arising from stresses. They non-reciprocally couple different components of the displacement field, leading either to waves with a linear dispersion even in non-inertial systems or to an instability with a growth rate that is linear in wavenumber~\cite{Rangan, AM_sol} in large enough systems. The interplay of chirality and uniaxiality (not necessarily polarity) allows for additional force densities that we shall see are equally important.

The force density that we will use in this article is
\begin{equation}
    \mathcal{F}^f_i = w_1(\Phi)p_i + w_2 C_{ij}p_j + w^c_1 \epsilon_{ikl} Q_{lm}C_{mk}.
    \label{frc}
\end{equation}
These potentially end up contributing terms at \emph{first} order in gradients in the uniaxial elastomeric phase. The first two of these~\cite{AM_sol} are achiral but polar, while the third is \emph{apolar} but chiral. Multiple (infinitely many) force densities that cannot be written as stresses are allowed that contribute terms at $\mathcal{O}(\nabla^2)$ in the uniaxial elastomeric or solid phase. These include force densities that are proportional to $Q_{ik}\partial_jQ_{kj}$, $Q_{ik}\partial_k\Phi$, $Q_{ik}\partial_jC_{kj}$ etc. We do not consider them here because they are only allowed in systems that do not conserve momentum and in that case, they are subdominant to the terms in~\eqref{frc}.

\section{Velocity-dependent force densities}
\label{sec:vforceden}

\numberwithin{equation}{section}
\renewcommand{\theequation}{\thesection.\arabic{equation}}

We now discuss velocity-dependent active and passive force densities. We start with the force densities that can be written as the divergence of a stress $\bm{\sigma}^v$ and are present even in systems that conserve momentum. This tensor can be written as the sum of an even $\bm{\sigma}^v_e$---viscous---and an odd $\bm{\sigma}^v_o$ part.
In equilibrium uniaxial materials, the viscosity tensor that enters $\bm{\sigma}^v_e=\eta_{ijkl}\partial_kv_l$ has the following form: 
\begin{multline}
\label{achivis}
    \eta_{ijkl} = \bar{\eta}_{1} p_i p_j p_k p_l + \bar{\eta}_2 \bigl( p_i p_j \bar{\delta}^T_{kl} + \bar{\delta}^T_{ij} p_k p_l \bigr) \\ 
    +(\bar{\eta}_3 - \bar{\eta}_4) \,\bar{\delta}^T_{ij}\bar{\delta}^T_{kl} + \bar{\eta}_4 \bigl( \bar{\delta}^T_{ik} \bar{\delta}^T_{jl} + \bar{\delta}^T_{il} \bar{\delta}^T_{jk} \bigr) \\ 
    + \frac{\bar{\eta}_5}{2} \bigl( p_i p_k \bar{\delta}^T_{jl} + p_i p_l \bar{\delta}^T_{jk} + p_j p_k \bar{\delta}^T_{il} + p_j p_l \bar{\delta}^T_{ik} \bigr) \,,
\end{multline}
where $\bar{\delta}^T_{ij}=p_kp_k\delta_{ij}-p_ip_j$. Note that all five viscosities are invariant under inversion, i.e. ${\bf p}\to -{\bf p}$, and $\eta_{ijkl}=\eta_{klij}=\eta_{jikl}=\eta_{ijlk}$. In a state with polar or uniaxial order with ${\bf p}=p_0{\bf n}^0$, and to zeroth order in fluctuations, we get
\begin{multline}
\label{etaijkl}
	\eta_{ijkl} = {\eta}_{1} n^0_i n^0_j n^0_k n^0_l + {\eta}_2 \bigl( n^0_i n^0_j {\delta}^T_{kl} + {\delta}^T_{ij} n^0_k n^0_l \bigr) \\
    + ({\eta}_3-{\eta}_4) \,{\delta}^T_{ij}\bar{\delta}^T_{kl} + {\eta}_4 \bigl( {\delta}^T_{ik} {\delta}^T_{jl} + {\delta}^T_{il} {\delta}^T_{jk} \bigr) \\
    + \frac{{\eta}_5}{2} \bigl( n^0_i n^0_k {\delta}^T_{jl} + n^0_i n^0_l {\delta}^T_{jk} + n^0_j n^0_k {\delta}^T_{il} + n^0_j n^0_l {\delta}^T_{ik} \bigr) \,,
\end{multline}
where ${\delta}^T_{ij}=\delta_{ij}-n^0_in^0_j$. We will often consider a simpler single viscosity limit obtained by taking $\eta_1=\eta_4=\eta_5=\eta$ and $\eta_2=\eta_3=0$. In chiral and polar materials, we have additional velocity-dependent active stresses at the same order in gradients~\cite{khain2022stokes, SJ_col}:
\begin{equation}
	\label{chivis0}
    	{\sigma_{ij}}^v_o = 2\left[\epsilon_{ilk}p_k\left\{2{{\eta}^c_2}p_jp_mA_{lm}+{\eta}^c_1\left(A_{lj}-p_jp_mA_{lm}\right)\right\}\right]^S\,,
\end{equation}
whose linearised form in a polar state ordered along ${\bf n}^0$ is given by
\begin{equation}
	\label{chivis}
	{\sigma_{ij}}^v_o=2\left[\epsilon_{ilk}n^0_k\left\{2{{\eta}^c_2}n^0_jn^0_mA_{lm}+{\eta}^c_1\left(A_{lj}-n^0_jn^0_mA_{lm}\right)\right\}\right]^S\,.
\end{equation}
This implies the force density
    \begin{equation}
    \label{chivisF}
    \boldsymbol{\mathcal{F}}^v_o = \eta^c_2 \nabla_\perp^2 \bm{\varepsilon} \cdot{\bf v}_\perp + {\eta^c_1} \partial_z \bigl( 2\Omega_{xy} \hat{\bf z} + \partial_z \bm{\varepsilon}\cdot{\bf v}_{\perp} +\bm{\varepsilon}\cdot\nabla_{\perp} v_z \bigr) , 
\end{equation}
when we take ${\bf n}^0\equiv\hat{{\bf z}}$ without loss of generality, with $\Omega_{xy}=(1/2)(\partial_x v_y-\partial_yv_x)$ \cite{AM_chi_rev, SJ_col}, $\perp\equiv(x,y)$, and
\begin{equation}
    \bm{\varepsilon}=\begin{pmatrix} 0& 1\\-1 &0\end{pmatrix}\,,
\end{equation} 
being the rank two Levi-Civita tensor. Note that unlike the viscosities in Eq.~\eqref{achivis}, the chiral and active velocity-dependent force densities in~\eqref{chivis} require polarity. 

When the gel moves through a fixed background medium, such that its momentum is not conserved, additional velocity-dependent force densities come into play. The simplest of these is the friction $\gamma{\bf v}$. We will set $\gamma=1$ throughout the article, without loss of generality. In addition to this, active and polar materials have other velocity-dependent force densities, the simplest of which is $\gamma_c\epsilon_{ijk}v_jp_k$. Amusingly, this has the same structure as the Lorentz force, with ${\bf p}$ playing the role of a magnetic field~\cite{AM_chi_rev}. However, as far as the linear theory is concerned, in the inertia-less limit, the effect of $\gamma_c$ is merely to renormalise the coefficients of the force densities in~\eqref{frc} as can be seen via the following argument: the ``friction tensor'' $\Gamma_{ij}$, which has an odd part, can be written as $\Gamma_{ij}=\delta_{ij}+\gamma_c\epsilon_{ijk}p_k$. The inverse of this is the mobility tensor $M_{ij}$, the odd part of which can also be seen to be $\propto\epsilon_{ijk}p_k$. Therefore, the additional chiral terms due to the presence of an odd velocity-dependent active force density are $\propto\epsilon_{ijk}p_k\mathcal{F}^f_j$ (to lowest order in gradients). It is easy to see that this does not yield any additional linear terms but merely shifts coefficients of those that are already present in $\bm{\mathcal{F}}^f$. Therefore, we will not discuss $\gamma_c$ further. 

\numberwithin{equation}{subsection}
\renewcommand{\theequation}{\thesection.\thesubsection.\arabic{equation}}

\section{Polar elastomers: passive force densities and relaxation of orientational fluctuations to displacement fluctuations}
\label{sec:passens}
In this section, we recapitulate the physics of passive polar elastomers and show that director fluctuations are completely determined by displacement fluctuations, implying that while polar elastomers spontaneously break rotation symmetry, the only hydrodynamic modes are the displacement fields. The rest of the section is organised as follows: in Sec.~\ref{elasen}, we show that even when elastomers break rotation symmetry spontaneously the fluctuations of the orientation direction are not soft; instead, they are fully determined by the local strains. In Sec.~\ref{instab_unifl} we show that, as a consequence of this, orientationally-ordered solids can be stable in chiral materials as well, though nematic and polar phases are linearly unstable in chiral passive and active fluids. In Sec.~\ref{subsec:elimact}, we show that orientational fluctuations can be eliminated in favour of strain fluctuations even in active chiral solids. The fluctuations about a uniaxial elastomeric state are best discussed in terms of an anisotropic strain tensor defined with respect to the uniaxial steady state. However, we built the theory about an isotropic reference state. In Sec.~\ref{flucan}, we follow~\cite{Ranjan_Lubensky} and show how to express fluctuations of all quantities---including the polarisation field---in terms of a strain tensor $\bm{E}$ defined with respect to a uniaxial reference. In Sec.~\ref{El_Fener} we write the free energy that would control the dynamics in the absence of activity, in terms of $\bm{E}$.

\subsection{Eliminating orientation fluctuations in favour of strain fluctuations in passive uniaxial elastomers}
\label{elasen}
\begin{figure}[t]
    \centering
    \includegraphics[width=\linewidth]{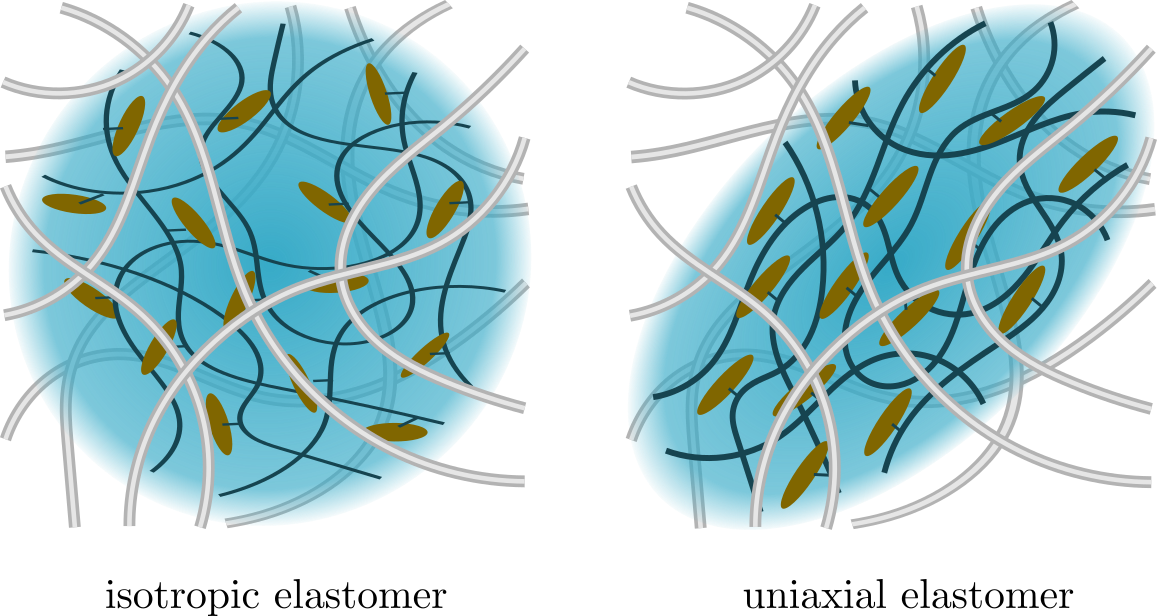}
    \caption{Schematic illustration of a uniaxial elastomer in an isotropic external medium. The left-hand panel shows the isotropic state (${\bf p} = {\bsf Q} = 0$) and the right-hand panel a spontaneously ordered state, as in Eq.~\eqref{strainss}. The elastomer is shown in blue with a side-chain polymer architecture; the external medium is visualised as a light grey mesh. Note that orientational ordering is associated with a deformation of the solid.} 
    \label{fig:elastomer}
\end{figure}

The phenomenological free energy that can describe the transition from an isotropic elastic state to a polar elastomer phase (see Fig.~\ref{fig:elastomer}) is given by
\begin{equation}
    F=\int d^3{\bf x} \left [\frac{\lambda}{2}\left(\Phi-\frac{s}{\lambda}p^2\right)^2+\mu\text{Tr}\left( {\bsf C}-\frac{t}{\mu}{\bsf Q}\right)^2+f_p \right ] \,,
    \label{FQ}
\end{equation}
where $f_p=-({r}/{2})p^2+({w}/{4})p^4$ and $\lambda, \mu$ are the bulk and shear moduli, respectively~\cite{Ranjan_Lubensky}. Chiral contributions arise at higher orders in gradients and fields than the ones retained here. The generic couplings with coefficients $s$ and $t$ between the strain and the polarisation lead to the development of a strain anisotropy either parallel ($t>0$) or perpendicular ($t<0$) to ${\bsf Q}$.
Taking a uniformly polarised state with ${\bf p} = p_0 \,\hat{\bf z}$, \eqref{FQ} implies 
\begin{equation}
    \Phi_0=\frac{sp_0^2}{\lambda} \,, \qquad {\bsf C}_0=\frac{t}{\mu}{\bsf Q}_0\,,
    \label{strainss}
\end{equation}
with $p_o = \sqrt{r/w}$. We consider the dynamics of linear fluctuations about this state. 

This calculation follows the one detailed in~\cite{Ranjan_Lubensky, AM_sol}. We summarise key details here and refer the interested reader to those works for a more detailed discussion. Expanding the free energy~\eqref{FQ} to harmonic order in fluctuations, we find that the fluctuations of ${\bf p}$, $\delta{\bf p}\equiv \delta{\bf p}_\perp + \hat{{\bf z}} \delta p_z$ are fully determined by those of the deviatoric and isotropic parts of the strain field, $\delta\bsf{C}$ and $\delta\Phi$: 
\begin{equation}
    \label{perpeq}
    \delta{\bf p}_\perp=\frac{\mu}{tp_0}\delta{\bsf C}_{z\perp}\,,
\end{equation}
and
\begin{equation}
\label{zeq}
    \delta p_z=\frac{6t\delta C_{zz}+3s\delta\Phi}{\bar{r}p_0}\,,
\end{equation}
where $\bar{r}=3w+6s^2/\lambda+8t^2/\mu$. The fluctuations $\delta{\bf p}_\perp$ would have a vanishing relaxation rate in the limit of zero wavenumber in polar \emph{fluids} that spontaneously break rotation symmetry but~\eqref{perpeq} implies that in polar \emph{elastomers} these fluctuations are not massless. This is a fundamental feature of systems that break both rotation and translation symmetry~\cite{Mermin} and can be understood in analogy with the Anderson-Higgs mechanism~\cite{Chaikin, Ranjan_Lubensky}. Importantly, this allows for the existence of polar or nematic \emph{elastomers} in chiral equilibrium systems, even though there is an obstacle to the existence of polar or nematic \textit{liquids} in chiral materials, as we show next. 

\subsection{Instability of uniaxial fluid in chiral systems}
\label{instab_unifl}
    
To see that polar or nematic \textit{liquids} are destabilised by chirality, recall that in uniaxial fluid states chirality contributes a term $f_{pc}=(k_c/2){\bf p}\cdot\nabla\times{\bf p}$ \cite{Oseen_LC,Frank_LC, deGen, lubensky1985hydrodynamics} to the free energy density. The passive dynamics of a polarisation vector is given by 
\begin{equation}
    \partial_t{\bf p}=-\frac{\delta F_{lc}}{\delta{\bf p}}\,,
\end{equation}
where $F_{lc}=\int d^3{\bf x}(f_{pc}+f_p)$. Then the linear dynamics of polarisation fluctuations $\delta{\bf p}_\perp$ transverse to the ordering direction $\hat{{\bf z}}$ about a polar (fluid) state ${\bf p}=\sqrt{r/w}\,\hat{{\bf z}}$ is given by
\begin{equation}
    \partial_t \delta{\bf p}_\perp = k_c \bm{\varepsilon} \cdot \partial_z \delta{\bf p}_\perp + \mathcal{O}\left(\nabla^2\right)\,,
\end{equation}
to the lowest order in gradients, where $\bm{\varepsilon}$ is the \emph{two-dimensional} Levi-Civita tensor. Thus, the eigenfrequency of polarisation fluctuations reads $\omega_\pm=\pm i k_cq_z$, where $q_z$ denotes the $z$-component of a wavevector ${\bf q}\equiv q_z\hat{{\bf z}}+{\bf q}_\perp$. This implies an instability of the polarised (or uniaxial) state with a growth rate linear in wavenumbers $q\equiv|{\bf q}|$. Indeed, the chiral contribution to the energy together with Frank elasticity that arises at next order in gradients leads to the formation of a helical, cholesteric state~\cite{deGen,Lubensy_chol}.

A similar instability affects a three-dimensional \emph{active} polar (or nematic) fluid state in chiral materials. To see this, consider the following equation for the polarisation field: 
\begin{equation}
\label{eq:polchi}
    \partial_t{\bf p}=(r-w |{\bf p}|^2){\bf p}+\lambda{\bf p}\cdot\nabla{\bf p}-\lambda_c{\bf p}\times{\bf p}\cdot\nabla{\bf p}-k_c\nabla\times{\bf p}+\mathcal{O}\left(\nabla^2\right)\,.
\end{equation}
The linearised dynamics of transverse fluctuations about the ordered state ${\bf p}=\sqrt{r/w}\,\hat{{\bf z}}$ in this case becomes
\begin{equation}
\label{eq:flucpolchi}
        \partial_t \delta{\bf p}_\perp = \lambda p_0 \partial_z \delta{\bf p}_\perp + (\lambda_cp_0^2 + k_c) \bm{\varepsilon} \cdot \partial_z \delta{\bf p}_\perp + \mathcal{O}\left(\nabla^2\right)\,.
\end{equation}
This implies an eigenfrequency  
\begin{equation}
\label{eq:poleigen}
    \omega_\pm=-q_z[\lambda p_0\pm i(\lambda_c p_0^2+k_c)]\,,
\end{equation}
which again indicates an instability with a linear growth rate. Importantly, in both passive and active fluids, the instability that destroys the uniaxial state in chiral matter has a \emph{linear} growth rate that vanishes at zero wavenumber. Our calculation of the polarisation fluctuations in passive elastomers implies that the coupling to elasticity leads to a \emph{finite relaxation rate} even for the transverse fluctuations of the polarisation field, even at zero wavenumber. This stabilises uniaxial order in elastomers at small wavenumbers and, depending on the values of Frank elastic coefficients, at all wavenumbers.

\subsection{Elimination of polarisation fluctuations in terms of displacement fluctuations even in active systems}
\label{subsec:elimact}
We derived the expression for polarisation fluctuations in terms of the strain variables in Eqs.~\eqref{perpeq} and~\eqref{zeq} from a free energy description. We now show that we can consistently use this expression to eliminate orientational fluctuations in favour of displacement fluctuations even in active systems. To do this, we consider an active model for coupled polarisation and velocity dynamics:
\begin{gather}
    \partial_t{p}_i=-\frac{\delta F}{\delta{p}_i}+\Lambda{v}_i+\alpha_c\epsilon_{ikl}Q_{lm}C_{mk}\,,
    \label{poleq} \\
    \Gamma{v}_i=\mathcal{F}^f_i-\Lambda\frac{\delta F}{\delta p_i}\,.
    \label{veqp}
\end{gather}
Here, we have only retained terms to the lowest order in gradients (which is why elastic force densities are not explicitly displayed) and $\bm{\mathcal{F}}^f$ was defined in Eq.~\eqref{frc}. Since the polarisation field is expected to have a finite relaxation rate at zero wavenumber, as in passive materials, the left-hand side of~\eqref{poleq} can be ignored at long-enough times. Then, we can solve for ${\bf v}$ by using~\eqref{veqp} and inject it into \eqref{poleq}. This yields
\begin{multline}
    \left(1+\frac{\Lambda^2}{\Gamma}\right) \frac{\delta F}{\delta{p}_i}=\frac{\Lambda w_1}{\Gamma}p_i+\frac{\Lambda w_2}{\Gamma}C_{ij}p_j\\+\left(\alpha_c+\frac{\Lambda w^c_1}{\Gamma}\right)\epsilon_{ikl}Q_{lm}C_{mk}\,.
    \label{polss}
\end{multline}
Similarly, the term $\delta F/\delta{\bf p}$ can be eliminated from~\eqref{veqp} by using the steady-state version of~\eqref{poleq} and redefining $\Gamma$ and the phenomenological coefficients in $\bm{\mathcal{F}}$. Now, the first two terms of the right-hand side of~\eqref{polss} can be absorbed into a redefinition of the coefficients in the free energy. Of course, the same redefinition would not have worked in~\eqref{veqp} but, as we have just argued, the free energy derivative in that equation can be eliminated by redefining the active force densities. This implies that the $\Lambda$ coupling cannot change the hydrodynamics of the \emph{polarised} elastomeric state (though, of course, it will change its onset and the values of phenomenological parameters such as elastic moduli in that state). 

Therefore, we now ignore the $\Lambda$ coupling and concentrate of the remaining part of~\eqref{polss}:
\begin{equation}
    \frac{\delta F}{\delta{p}_i}=\alpha_c\epsilon_{ikl}Q_{lm}C_{mk}\,.
    \label{polss2}
\end{equation}
Now, we see that the steady state we discussed earlier, given by ${\bf p}=\sqrt{r/w}\,\hat{{\bf z}}$, and~\eqref{strainss}, is still a steady state of~\eqref{polss2}. The reason is that the right-hand side of~\eqref{polss2} vanishes in this case: ${\bsf C}_0\propto{\bsf Q}_0$ and is a diagonal matrix. Therefore, ${\bsf C}_0\cdot{\bsf Q}_0$ is also diagonal, and the full contraction of that with a totally antisymmetric tensor vanishes. We now show that~\eqref{perpeq} and~\eqref{zeq}, implicitly obtained by setting $\delta F/\delta{\bf p}=0$ at first order in fluctuations, remains correct to linear order in fluctuations even for~\eqref{polss2}. To do this, we expand the right-hand side of~\eqref{polss2} and show that it vanishes when we use~\eqref{perpeq} and~\eqref{zeq}. To linear order in fluctuations,
\begin{multline}
    \epsilon_{ijk}Q_{kl}C_{jl} \approx \epsilon_{ijk} (\delta Q_{kl} C_{0_{jl}} + Q_{0_{kl}} \delta C_{jl}) \\
    = \epsilon_{ijk} \left(\frac{t}{\mu} \delta Q_{kl} Q_{0_{jl}} + Q_{0_{kl}} \delta C_{jl} \right)\,.
\label{lineqfl}
\end{multline}
Due to the Levi-Civita tensor, only terms with $z\perp$ components of the fluctuating field can contribute. Eq.~\eqref{perpeq} implies that $\delta {\bsf Q}_{z\perp}=(\mu/t)\delta{\bsf C}_{z\perp}$. Inserting this in~\eqref{lineqfl}, we see that it vanishes because the tensor in the brackets in~\eqref{lineqfl}, that is fully contracted with the Levi-Civita tensor, is symmetric. Thus, we can continue using $\delta F/\delta{\bf p}=0$ to eliminate polarisation fluctuations to strain fluctuations even in chiral and active systems.

\subsection{Fluctuations about the anisotropic state}
\label{flucan}
The steady state of polar elastomers is anisotropic. Indeed, Eq.~\eqref{strainss} shows that ${\bsf U}_0={\bsf C}_0+(1/3)\Phi_0{\bsf I}$---which is a strain tensor defined with respect to a fiducial isotropic state---is non-zero in this steady state. While we can carry out calculations of fluctuations about the anisotropic state using this strain tensor and the displacement field about an isotropic state, it is much more convenient to use a displacement field and a strain tensor with respect to the anisotropic state. To this end, we define a displacement field ${\bf u}$ and a displacement gradient field $\bm{E}\equiv\nabla{\bf u}$ with respect to the anisotropic reference state. To obtain the transformation from ${\bsf U}$ to $\bm{E}$, we first define the steady-state deformation tensor of the anisotropic state:
\begin{equation}
    \boldsymbol{\Lambda}_0\boldsymbol{\Lambda}_0^T=2{\bsf U}^0+{\bsf I}\,,
\end{equation}
where $\boldsymbol{\Lambda}_0$ is the deformation matrix of the homogeneously polarised state, which we write as 
\begin{equation}
    \boldsymbol{\Lambda}_0 \boldsymbol{\Lambda}_0^T = \begin{pmatrix} \Lambda^2_\perp & 0 & 0 \\ 0 & \Lambda^2_\perp & 0 \\ 0 & 0 & \Lambda^2_\parallel \end{pmatrix}\,,
\end{equation}
with
\begin{equation}
\label{lampar}
    \Lambda^2_\parallel=\frac{2p_0^2}{3}\left(\frac{s}{\lambda}+\frac{2t}{\mu}\right)+1\,,
\end{equation}
and 
\begin{equation}
\label{lamperp}
    \Lambda^2_\perp=\frac{2p_0^2}{3}\left(\frac{s}{\lambda}-\frac{t}{\mu}\right)+1\,.
\end{equation}
The anisotropy is  
\begin{equation}
\label{lamaniso}
    \Lambda^2_\parallel-\Lambda^2_\perp=\frac{2t}{\mu}p_0^2\,.
\end{equation}
We also define $\mathcal{R}={\Lambda^2_\parallel}/{\Lambda^2_\perp}$. The deviations about the polarised state are characterised by the strain tensor
\begin{equation}
    \delta{\bsf U}=\frac{1}{2}(\boldsymbol{\Lambda}\boldsymbol{\Lambda}^T-\boldsymbol{\Lambda}_0\boldsymbol{\Lambda}_0^T)\,.
\end{equation}
We further define the strain tensor ${E}^S_{ij}=(\partial_iu_j+\partial_ju_i)/2$ and the antisymmetric displacement gradient tensor ${E}^A_{ij}=(\partial_iu_j-\partial_ju_i)/2$, in terms of which, the components of $\delta{\bsf U}$ become
\begin{gather}
\label{Wppu}
    \delta\bsf{U}_{\perp\perp}=\Lambda_\perp^2\bm{E}^S_{\perp\perp}\,, \\
\label{Wzpu}
    \delta\bsf{U}_{\perp z}=\delta\bsf{U}_{z\perp}=(\beta^{-1}\bm{E}^S_{z\perp}+\bm{E}^A_{z\perp})\frac{\Lambda^2_\parallel-\Lambda^2_\perp}{2}\,, \\
\label{Wzzu}
    \delta\bsf{U}_{zz}=\Lambda_\parallel^2\bm{E}^S_{zz}\,.
\end{gather}
These are the transformation rules used in constructing the hydrodynamic theory of a uniaxial elastomer in terms of the ${\bf u}$ field.
Using these, we express Eq.~\eqref{perpeq} for polarisation fluctuations in terms of $\bm{E}^S$ and $\bm{E}^A$ as~\cite{Olmsted, Warner_Terentjev}
\begin{equation}
\label{pperpu}
    \delta{\bf p}_\perp=p_0(\beta^{-1}\bm{E}^S_{z\perp}+\bm{E}^A_{z\perp})\,,
\end{equation}
where 
\begin{equation}
    \beta=\frac{\mathcal{R}-1}{\mathcal{R}+1}=\frac{\Lambda_\parallel^2-\Lambda_\perp^2}{\Lambda_\parallel^2+\Lambda_\perp^2}\,.
\end{equation}

\subsection{Elastomeric free energy in terms of $\boldsymbol{E}$}
\label{El_Fener}
In this subsection, we express the free energy in~\eqref{FQ} in terms of $\bm{E}$ using~\eqref{Wppu},~\eqref{Wzpu},~\eqref{Wzzu} and~\eqref{pperpu}. Expanding~\eqref{FQ} to harmonic order in fluctuations and integrating out $\delta p_z$ fluctuations, we get
\begin{multline}
    \label{FU}
	F=\frac{1}{2}\int d^3{\bf x}\Big\{B_1(E^S_{zz})^2+2B_2E^S_{zz}\left(E^S_{xx}+E^S_{yy}\right)\\+B_3\left(E^S_{xx}+E^S_{yy}\right)^2+2B_4\left({E^S_{xx}}^2+{E^S_{yy}}^2+2{E^S_{xy}}^2\right)\\+\bar{\mu} \bigl[ \beta(\delta{\bf p}_\perp/p_0-\bm{E}^A_{z\perp})-\bm{E}^S_{z\perp} \bigr]^2 \Big\}\,,
\end{multline}
where $\bar{\mu}$ is related to $\mu$ in~\eqref{FQ} as $\bar{\mu}=\mu(\Lambda_\parallel^2+\Lambda_\perp^2)^2$.

The relations between $B_{1-4}$ and the coefficients in~\eqref{FQ} are not very interesting; they are displayed in~\cite{AM_sol} and we don't present them here for brevity. Importantly, it is immediately clear that integrating out the massive $\delta{\bf p}_\perp$ fluctuations also eliminates any cost for $\bm{E}^S_{z\perp}$ shears. The free energy after integrating out $\delta{\bf p}_\perp$ reads
\begin{multline}
    \label{FU2}
	F = \frac{1}{2} \int d^3{\bf x} \Big[ B_1 (E^S_{zz})^2 + 2B_2 E^S_{zz} \left(E^S_{xx}+E^S_{yy}\right) \\
    + B_3 \left(E^S_{xx}+E^S_{yy}\right)^2 + 2B_4 \left({E^S_{xx}}^2+{E^S_{yy}}^2+2{E^S_{xy}}^2\right) \Big]\,,
\end{multline}    
which implies that, in equilibrium polar (or uniaxial) elastomers, there is no force density $\propto\nabla_\perp^2u_z \,\hat{{\bf z}}$ or $\partial_z^2{\bf u}_\perp$. This fact makes $u_z$ fluctuations of an equilibrium elastomer analogous to the fluctuations of the displacement field in smectics and the ${\bf u}_\perp$ fluctuations analogous to displacement fluctuations in columnar materials.

The linearised elastic force densities are given by $-\delta F/\delta{\bf u}$, where $F$ is the energy in~\eqref{FU2}. These force densities can also be written as divergences of stresses: $-\delta F/\delta{\bf u}=\nabla\cdot\bm{\sigma}^e$.

\section{The effect of chiral activity on polar elastomers}
\label{sec:chielexp}

\begin{figure*}[t]
    \centering
    \includegraphics[width=0.95\linewidth]{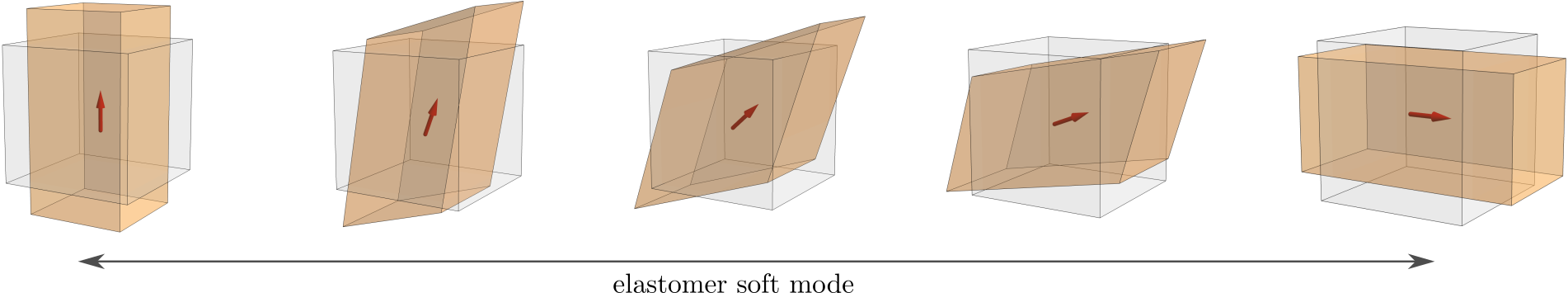}
    \caption{Illustration of the soft mode of elastomers. The isotropic reference state is indicated as a light grey cube and the uniaxial state as an orange parallelepiped with the direction of polar alignment indicated by a red arrow. These are all equivalent ground states of the elastomer, Eq.~\eqref{strainss}, but corresponding to different elastic strains of the isotropic state. Importantly, the intermediate states cannot be obtained via a rigid rotation of the leftmost parallelepiped. This is the key distinction between the symmetry properties of uniaxial elastomers and solids: the latter are invariant only under rigid rotations of the parallelepiped.}
    \label{fig:softmode}
\end{figure*}

In this section, we examine the effect of chiral activity on polar elastomers. We start in Sec.~\ref{subsec:chifor} by examining the effect of chiral and active force densities discussed in Sec.~\ref{frcactfric}, which arise in the presence of a fixed momentum sink, on linear fluctuations of uniaxial elastomers. Surprisingly, we show that there is no effect because of the somewhat unique elastomeric symmetry. In Sec.~\ref{subsec:achioddel} we present a discussion of chiral and achiral odd elasticities in the context of active polar elastomers. In Sec.~\ref{Dynregime}, we schematically display the equations of motion for the velocity and the displacement fields that we will use to construct the dynamics of the displacement field in three distinct physical regimes. In Sec.~\ref{elfixback}, we construct the dynamics of the displacement field for active, chiral and polar elastomers moving through a fixed background medium. In Sec.~\ref{elasfl} we consider the case in which the medium is a momentum-conserved, Stokesian fluid. Finally, in Sec.~\ref{elasin}, we examine the limit in which there is no dissipation at all.

\subsection{Chiral force densities in elastomers in a fixed background medium}
\label{subsec:chifor}
The force density $\bm{\mathcal{F}}^f$ discussed in~\eqref{frc} can be evaluated in a polar elastomeric state using the results summarised in Secs.~\ref{elasen} and~\ref{flucan}. We note that the first two terms, which are achiral, imply that the solid has a non-zero speed
\begin{equation}
    v_0=\left[w_1(\Phi_0)+\frac{2w_2t}{3\mu}p_0^2\right]p_0\,,
    \label{speed}
\end{equation}
in the steady state with respect to the fixed background medium, where $p_0=\sqrt{r/w}$. The chiral term vanishes in the steady state as can be seen from the discussion below~\eqref{polss2}.

We now expand Eq.~\eqref{frc} to first order in fluctuations and use the results in Secs.~\ref{elasen} and~\ref{flucan} to convert them to expressions in terms of ${\bf u}$. For the achiral terms, this follows~\cite{AM_sol} with the same result, which we do not reiterate here. We next show that the chiral term in~\eqref{frc} vanishes to leading order in fluctuations in a polar elastomeric state. Expanding the chiral active force density to linear order in fluctuations, we obtain
\begin{equation}
\label{apolfluc}
    w^c_1\epsilon_{ijk}Q_{jl}C_{kl}\approx w^c_1\epsilon_{ijk}(\delta Q_{kl}C_{0_{jl}}+Q_{0_{kl}}\delta C_{jl})\,.
\end{equation}
Now, the only non-zero contribution to this can come from the $z\perp$ components of the fluctuating fields. However, from~\eqref{strainss}, ${\bsf Q}_0=(t/\mu){\bsf C}_0$ and~\eqref{perpeq} implies that the fluctuations $\delta {\bsf Q}_{z\perp}=(\mu/t)\delta{\bsf C}_{z\perp}$. Therefore, the tensor fully contracted with $\bm{\epsilon}$ is symmetric, and therefore, this term vanishes. 
 
Note that the chiral active force density in~\eqref{frc} is invariant under ${\bf p}\to-{\bf p}$, i.e., it is allowed in apolar but uniaxial active materials as well. Further, the argument in the paragraph above works for a nematic elastomer as well and shows that its linear dynamics in a frictional background medium is insensitive to chirality to leading order in gradients. In the next section, we will show that chiral activity does not affect even a momentum-conserved nematic elastomer. 

The reason that the chiral active force density does not affect the linear dynamics of a uniaxial elastomeric phase is that the fluctuations of the strain tensor $\delta{\bsf C}_{z\perp}$ and that of the order parameter $\delta{\bsf Q}_{z\perp}$ are proportional to each other. This happens because a uniform director rotation in a nematic elastomer can be compensated by a strain, as illustrated in Fig.~\ref{fig:softmode}. The alignment direction is arbitrary but dictates the anisotropy of the steady-state strain; therefore, rotating the director without a countervailing rotation of the elastic medium, which yields an elastically distinct state with a rotated anisotropy of the steady-state strain, is a symmetry. This invariance is in contrast with that of polar or nematic liquid crystals, which are symmetric only under \emph{joint} rotations of the anisotropy direction and the coordinates. As we will discuss later, this invariance, which distinguishes uniaxial elastomers from uniaxial solids that are only invariant under joint rotations of the director and the elastic medium, is broken in physical materials. In passive systems, the higher symmetry of uniaxial elastomers eliminates one of the elastic moduli that characterises uniaxial solids, as we discussed in Sec.~\ref{El_Fener} and makes them much less resistant to long-wavelength fluctuations. In fact, passive uniaxial elastomers only have quasi-long range order in three dimensions.

In~\cite{AM_sol}, some of us showed that the linear theory of {active} \emph{achiral} uniaxial elastomers was analogous to {passive} or active uniaxial \emph{solids}: activity led to force densities in uniaxial elastomers that were forbidden by symmetry in passive elastomers but not in solids. In other words, the greater symmetry of the uniaxial but achiral elastomer did not affect the hydrodynamics of these materials in the presence of activity. Our calculation here, taken together with the discussion of the effect of activity on chiral uniaxial solids in the accompanying Letter~\cite{short} and later in this article in Sec.~\ref{sec:Unisol}, shows that the additional symmetry of elastomers makes a dramatic difference in chiral materials. 

Unlike the chiral force density, the contribution of achiral but polar force densities with coefficients $w_1$ and $w_2$ does not vanish at linear order in fluctuations. These contributions, which appear at $\mathcal{O}(\nabla)$ and therefore dominate over elastic force densities, were discussed in detail in~\cite{AM_sol, AM_sol_err}; we will not repeat the calculation here but simply state the result in Sec.~\ref{elfixback}.

\subsection{Chiral and achiral odd elasticity}
\label{subsec:achioddel}
In this section, we expand the chiral and active stress in~\eqref{cstrs} to linear order in fluctuations, using the results in Secs.~\ref{elasen} and~\ref{flucan}. Using~\eqref{strainss}, it can be checked that $\bm{\sigma}^{ac}$ vanishes in the steady state. We now expand each term of $\bm{\sigma}^{ac}$ to first order in fluctuating fields. First considering the term with the coefficient $\zeta^c_1$, using the strain tensor $\bsf{C}$ defined about the isotropic state,  we obtain
\begin{multline}
    4 \zeta^c_1(\epsilon_{ilk} p_k p_j p_m C_{lm})^S = 4\zeta^c_1 [p_j(\epsilon_{ilk}p_k p_m C_{lm})]^S \\= 4\zeta^c_1 [p_j(\epsilon_{ilk}Q_{km}C_{lm})]^S\,,
\end{multline}
where for the second equality, we have used the fact that $\epsilon_{ilk}\delta_{km}C_{lm}=\epsilon_{ilk}C_{lk}=0$. Here the final equality follows from the fact that $\bsf{C}$ is a symmetric tensor. The vector in round brackets, $\epsilon_{ilk}Q_{km}C_{lm}$ is the one we discussed in the last section---see Eq.~\eqref{apolfluc}. As we discussed there, it vanishes both in the steady state and to first order in fluctuations. This implies that there is no contribution $\propto\zeta^c_1$ to the linearised force density. Indeed, directly expanding the stress term to the first order in fluctuations,
\begin{widetext}
\begin{multline}
    4\zeta^c_1 (\epsilon_{ilk} p_k p_j p_m C_{lm})^S \approx 2\zeta^c_1 p_0^2 \begin{pmatrix} 0 & 0 & p_0\delta C_{yz}-C_{0_{zz}}\delta p_y \\ 0 & 0 & -p_0\delta C_{xz}+C_{0_{zz}}\delta p_x \\ \delta C_{yz}-C_{0_{zz}}\delta p_y & -p_0\delta C_{xz}+C_{0_{zz}}\delta p_x & 0 \end{pmatrix} \\ = 2\zeta^c_1 p_0^3 \begin{pmatrix} 0 & 0 & \delta C_{yz}-\frac{tp_0}{\mu}\delta p_y \\ 0 & 0 & -\delta C_{xz}+\frac{tp_0}{\mu}\delta p_x \\ \delta C_{yz}-\frac{tp_0}{\mu}\delta p_y & -\delta C_{xz}+\frac{tp_0}{\mu}\delta p_x & 0 \end{pmatrix}\,,
\label{vanoddstress}
\end{multline}
\end{widetext}
where we have used~\eqref{strainss} for the second equality. Eq.~\eqref{perpeq} then implies that~\eqref{vanoddstress} vanishes.

In contrast, the stress term $\propto\zeta^c_2$ \emph{does not} vanish to first order in fluctuations. Expanding this term to first order in fluctuations, simplifying using the fact that $(\epsilon_{ilk}p_kp_jp_mC_{lm})^S$ vanishes to first order in fluctuations, and using~\eqref{strainss} and~\eqref{perpeq}, we obtain
\begin{multline}
    2|{\bf p}|^2\left(\epsilon_{ilk}p_kC_{lj}\right)^S\approx p_0^3\begin{pmatrix} 2\delta C_{xy} & \delta C_{yy}-\delta C_{xx} &0\\\delta C_{yy}-\delta C_{xx} &-2\delta C_{xy} &0\\0 &0 &0\end{pmatrix}\\=p_0^3\Lambda_\perp^2\begin{pmatrix} 2E_{xy} & E_{yy}-E_{xx} &0\\E_{yy}-E_{xx} &-2E_{xy} &0\\0 &0 &0\end{pmatrix}\,,
\end{multline}
where for the last equality, we use~\eqref{Wppu}. This odd stress implies a force density $\propto\, \zeta^c_2 \nabla_\perp^2 \bm{\varepsilon} \cdot {\bf u}_\perp$. This term is analogous to the two-dimensional odd elasticity first discussed in~\cite{Echebarria1, Echebarria2, Madruga} and established as a generic consequence of chiral activity in~\cite{scheibner2020odd}; see also~\cite{fruchart2023odd, AM_chi_rev}. Here, it appears in a three-dimensional material in the directions transverse to the polarisation. This is similar to the appearance of the odd elasticity in the polar columnar phase~\cite{SJ_col}. 

Interestingly, our calculation here explicitly demonstrates that, just like in the columnar phase~\cite{SJ_col}, the interplay of activity and chirality yields only one chiral odd elastic term in the polar elastomeric phase as well. In the context of the columnar phase, this was attributed to the fact that it has a two-dimensional displacement field. However, despite the elastomer having a three-dimensional displacement field, the invariance under independent rotation of the solid structure and ordering direction~\cite{Olmsted} allows for only one chiral odd elastic force density rather than the two that appear in chiral polar solids we discuss in Sec.~\ref{sec:Unisol}; see also~\cite{short}.

Note that, as mentioned in Sec.~\ref{subsec:active_elastic_stresses}, the chiral odd force densities \emph{require} polarity. That is, even $\zeta^c_2$ is absent in chiral active nematic elastomers. 

As in the last section, we do not display the calculation for the achiral active stress in detail, as this was discussed in~\cite{AM_sol}. However, we highlight a few key points. The achiral active stress $\zeta_1{\bsf Q}$ in~\eqref{actacstrs} can be expanded to linear order in fluctuations using~\eqref{perpeq} and~\eqref{zeq}, and written in terms of the displacement field ${\bf u}$ using~\eqref{Wppu},~\eqref{Wzpu} and~\eqref{Wzzu}. The $z\perp$ components of this stress are the most interesting, and we discuss these:
\begin{multline}
    \zeta_1\delta{\bsf Q}_{z\perp}\approx\frac{\zeta_1\mu}{t}\delta{\bsf C}_{z\perp}\\=\frac{\zeta_1\mu(\Lambda_\parallel^2-\Lambda_\perp^2)}{2t}(\beta^{-1}\bm{E}^S_{z\perp}+\bm{E}^A_{z\perp})\\=\zeta_1p_0^2(\beta^{-1}\bm{E}^S_{z\perp}+\bm{E}^A_{z\perp})\,,
    \label{passunistrss}
\end{multline}
where the first approximate equality uses~\eqref{perpeq}, the first equality uses~\eqref{Wzpu} and the second equality uses~\eqref{lamaniso}. There are two interesting aspects of this stress. First, as discussed in detail in~\cite{AM_sol}, it implies force densities $\propto\nabla_\perp^2u_z\hat{{\bf z}}$ and $\partial_z^2{\bf u}_\perp$ which are forbidden in equilibrium systems due to a combination of rotation symmetry and time-reversal symmetry. Second, again as noted in~\cite{AM_sol}, the stress depends on $\bm{E}^A_{z\perp}$. This implies the existence of an achiral odd elasticity in active three-dimensional uniaxial (not necessarily polar) materials. This achiral odd elasticity was first pointed out by~\cite{old_odd} and was recently discussed by~\cite{shankar2024active}. That~\eqref{passunistrss} must violate the Maxwell-Betti relation can be understood from the following argument: due to the presence of $\bm{E}^A_{z\perp}$, the linearised stress tensor cannot be written as 
\begin{equation}
    \sigma_{ij}=W_{ijkl}\partial_ku_l\,,
\end{equation}
with $W_{ijkl}$ being symmetric under the interchanges $i\leftrightarrow j$, $k\leftrightarrow l$ and $\{i,j\}\leftrightarrow \{k,l\}$. Indeed, a generic stress-strain relation that is achiral but violates this condition can be written in terms of the polarisation vector as 
\begin{equation}
    \sigma^{on}_{ij}\propto(p_ip_j\delta_{kl}-\delta_{ij}p_kp_l)\partial_ku_l\,,
    \label{noreciachi}
\end{equation}
which is invariant under ${\bf p}\to-{\bf p}$ and can appear in apolar phases as well. To linear order in fluctuations, this stress implies a force density $\propto\hat{{\bf z}}\,\partial_z\nabla\cdot{\bf u}-\nabla\partial_zu_z$. The force density that arises due to the $\bm{E}^A$ dependence of~\eqref{passunistrss} is $\propto [\partial_z(\nabla_\perp u_z-\hat{{\bf z}}\nabla_\perp\cdot{\bf u}_\perp)+(\hat{{\bf z}}\nabla_\perp^2u_z-\partial_z^2{\bf u}_\perp)]$. The first part of this force density is exactly the one obtained from~\eqref{noreciachi} implying that the uniaxial active stress generically generates an achiral odd elasticity.

\subsection{Equation for the velocity field and the displacement field}
\label{Dynregime}
In this section, we construct the dynamical equations for the displacement fields ${\bf u}$ about the elastomeric state. When momentum is conserved, the momentum density is ultimately a slow field. However, if we confine ourselves to the length and timescales---relevant for most biological and soft matter systems---for which inertia is irrelevant, the momentum density is not a hydro\emph{dynamic} field. Instead, it is given by a force balance equation which, for momentum-conserved systems, takes the form $\nabla\cdot\bm{\sigma}^\text{tot}=0$, where $\bm{\sigma}^\text{tot}=\bm{\sigma}^a+\bm{\sigma}^v+\bm{\sigma}^e-\Pi{\bsf I}$ with $\Pi$ being the pressure which enforces the incompressibility constraint in gels and yields a \emph{constitutive} equation for the velocity field. Then using the linear relation $\partial_t{{\bf u}}={\bf v}$, we obtain a linearised dynamical equation for ${\bf u}$.

For the sake of completeness, we will also discuss the inertial case, though it is less relevant for biological materials. In this case, the linearised dynamical equation for the momentum density is given by 
\begin{equation}
\label{inertial}
    \rho_0\partial_t{\bf v}=\rho_0\partial_t^2{\bf u}=\nabla\cdot\bm{\sigma}^\text{tot}\,.
\end{equation}
For this case, we will explicitly construct the equations of motion only in the limit of zero viscosity and no velocity-dependent active stresses, i.e., no odd viscosity. 

When momentum is not conserved, its density is not a hydrodynamic variable. In this case, the linearised force balance equation reads 
\begin{equation}
    {\bf v}=\bm{\mathcal{F}}^f-\frac{\delta F}{\delta{\bf u}}+\nabla\cdot\bm{\sigma}^a\,,
\end{equation}
where, without loss of generality, we have set the friction coefficient to $1$, not included the terms coming from $\bm{\sigma}^v$---since they are subdominant to the frictional force density---and not included any other velocity-dependent chiral force densities for reasons discussed in Sec.~\ref{sec:vforceden}.

In permanently crosslinked systems, such as the one we consider, the displacement field dynamics is completely determined by the velocity field. In Eulerian coordinates, its dynamics reads
\begin{equation}
    \partial_t{\bf u}+{\bf v}\cdot\nabla{\bf u}={\bf v}\,.
\end{equation}
In momentum-conserved systems, the advective term is nonlinear, but for solids and elastomers moving through a fixed background medium, the mean value of $v_z=v_0\neq 0$. This implies that $\partial_tu_z$ has a non-zero mean value and that the advective term has a linear contribution $v_0\partial_z {\bf u}$ to the displacement dynamics. We eliminate the former via a redefinition $u_z\to u_z-v_0t$ and the latter via a Galilean shift of the $z$ coordinate $z\to z-v_0t$. 

We now combine the relations discussed in this section with the force densities discussed in the earlier sections to finally construct the linear hydrodynamic equations for the displacement fields for chiral active elastomers in various physical situations and dynamical limits.

\subsection{Dynamical equations for active polar and chiral elastomers in a fixed, external background medium}
\label{elfixback}
We construct the dynamical equations for the displacement field for a polar chiral elastomer in the case in which the background medium is fixed, implying that a frictional force balances other forces. 
As discussed earlier, all effects of chirality vanish at leading order in gradients in this case with the displacement field fluctuations about the steadily-moving state reading
\begin{gather}
    \label{elasuz}
    \partial_tu_z=a_1\partial_z u_z+a_2\nabla_\perp\cdot{\bf u}_\perp\,, \\
    \label{elasup}
    \partial_t{\bf u}_\perp= a_3\partial_z{\bf u}_\perp+a_4\nabla_\perp u_z\,.
\end{gather}
The various coefficients that appear in these equations can be related to the ones in~\eqref{frc} using~\eqref{perpeq}, \eqref{zeq}, \eqref{lampar}, \eqref{lamperp}, \eqref{lamaniso}, \eqref{Wppu}, \eqref{Wzpu} and~\eqref{Wzzu}. This yields
\begin{gather}
    \begin{split}
            a_1 & = \frac{\Lambda_\parallel^2}{3\bar{r}p_0\mu} \Bigl[ 3(3s+4t) w_1 \mu \\
            & \quad + p_0^2 \bigl( 6st w_2 + 8t^2 w_2 + 2\bar{r} w_2 \mu + 3\bar{r} w_3 \mu \bigr) \Bigr]\,,
    \label{eq:a1}
    \end{split} \\
    \begin{split}
            a_2& = \frac{\Lambda_\perp^2}{3\bar{r}p_0\mu} \Bigl[ 3(3s-2t) w_1 \mu \\
            & \quad + p_0^2 \bigl( 6st w_2 - 4t^2 w_2 - \bar{r} w_2 \mu + 3\bar{r} w_3 \mu \bigr) \Bigr] \,,
    \label{eq:a2}
    \end{split} \\
    a_3=(v_0/2)(\beta^{-1}+1)\,,\label{eq:a3}\\
    a_4=(v_0/2)(\beta^{-1}-1)\,.\label{eq:a4}
\end{gather}
However, it is more useful to consider $a_i$ as phenomenological parameters that need to be measured in experiments. In particular, any relation that $a_i$ may have here is not universal \emph{except} for the one connecting $a_3$ and $a_4$: $a_3/a_4=\mathcal{R}$, where $\mathcal{R}$ is defined below~\eqref{lamaniso}; this is required by rotation symmetry. 

To the lowest order in gradients $\mathcal{O}(\nabla)$, these equations have the same structure as those for externally-driven solids~\cite{Rangan} (the symmetry-mandated relation between $a_3$ and $a_4$ doesn't affect the scaling of displacement fluctuations~\cite{AMCMSR}). The leading-order behaviour implied by~\eqref{elasuz} and~\eqref{elasup} has been discussed in detail elsewhere~\cite{Rangan, AM_sol, AMCMSR}. Briefly, they imply the dispersion relation $\omega_\pm={1}/{2}\left[(a_1+a_3)q_z\pm\sqrt{(a_1-a_3)^2q_z^2+4a_2a_4q_\perp^2}\right]$ which yields a wavelike mode when $a_2a_4>0$ (with the wave speed vanishing along the direction $q_\perp^2/q_z^2=a_1a_3/a_2a_4$ when $a_1a_3>0$) even in this inertialess material and an unstable mode with growth rate $\propto |q_\perp|$ for $a_2a_4<0$. The limit $a_2a_4=0$, in which either ${\bf u}_\perp$ decouples from the $u_z$  dynamics or vice versa, is an exceptional point; the dynamics at precisely this point is also of interest and has been examined in~\cite{Das_Barma}.

\subsection{Dynamical equations for active polar and chiral elastomers in a fluid medium}
\label{elasfl}
In this section, we display the dynamical equations for a polar and chiral elastomeric gel in a Stokesian fluid, to leading order in gradients. The elastomer and the fluid together conserve momentum. Thus, the linearised force balance equation, in this case, is 
\begin{equation}
    \boldsymbol\eta \,\smvdots\, \nabla\nabla{\bf v}+ \boldsymbol{\mathcal{F}}^v_o=\nabla\Pi-\nabla\cdot\bm{\sigma}^a+\frac{\delta F}{\delta{{\bf u}}},
    \label{fbal}
\end{equation}
where $\boldsymbol{\eta}$ is the rank-4 tensor defined in~\eqref{etaijkl}, $\Pi$ is the pressure that enforces the three-dimensional incompressibility constraint $\nabla\cdot{\bf v}=0$, $\boldsymbol{\mathcal{F}}^v_o$ is displayed in~\eqref{chivisF}, $\bm{\sigma}^a$ is the sum of $\bm{\sigma}^{anc}$ and $\bm{\sigma}^{ac}$ displayed in~\eqref{actacstrs} and~\eqref{cstrs} respectively, expanded to linear order in fluctuations and expressed in terms of ${\bf u}$, and $F$ is the free energy in~\eqref{FU2}. Note that since the total momentum is conserved in this case, force densities must be divergences of a stress. Therefore, the force density $\bm{\mathcal{F}}^f$ in~\eqref{frc} does not appear in this equation. Using the results of Sec.~\ref{sec:passens} and~\ref{sec:chielexp} we write~\eqref{fbal} explicitly in terms of ${\bf u}$: 
\begin{widetext}
  \begin{equation}
        0 = -\partial_z \Pi + b_{\zeta1} \nabla_\perp^2 {u}_z + \eta_5 \nabla_\perp^2 {v}_z + (b_1 \partial_z^2 u_z + \eta_1 \partial_z^2 v_z) + b_2 \partial_z \nabla_\perp \cdot {\bf u}_\perp + \eta_2 \partial_z \nabla_\perp \cdot {\bf v}_\perp + \eta^c_1 \partial_z \nabla_\perp \times {\bf v}_\perp \,,
        \label{zvel1}
  \end{equation}
  \begin{multline}
        0= -\nabla_\perp\Pi+b_{\zeta2}\partial_z^2{\bf u}_\perp+\eta_5\partial_z^2{\bf v}_\perp+b_3\nabla_\perp\nabla_\perp\cdot{\bf u}_\perp+\eta_3\nabla_\perp\nabla_\perp\cdot{\bf v}_\perp+b_4\nabla_\perp^2{\bf u}_\perp+\eta_4\nabla_\perp^2{\bf v}_\perp+b_5\nabla_\perp\partial_zu_z\\+\eta_2\nabla_\perp\partial_zv_z+b^c_2\nabla_\perp^2\bm{\varepsilon}\cdot{\bf u}_\perp+\eta^c_2\nabla_\perp^2\bm{\varepsilon}\cdot{\bf v}_\perp+\eta^c_1\partial_z\bm{\varepsilon}\cdot(\partial_z{\bf v}_\perp+\nabla_\perp{v}_z)\,.
        \label{perpvel1}
  \end{multline}
\end{widetext}
While the coefficients in Eqs.~\eqref{zvel1} and~\eqref{perpvel1} should be taken as phenomenological parameters in their own right, whose values should be determined directly from experiments, we display their relations to the coefficients in~\eqref{actacstrs}, \eqref{cstrs}, \eqref{zeq}, \eqref{perpeq}, \eqref{Wppu}, \eqref{Wzpu}, \eqref{Wzzu} and~\eqref{FU2}:
\begin{equation}
    \label{eq:bc}
    b^c_2=p_0^3\zeta^c_2\Lambda_\perp^2\,, 
\end{equation}
\begin{equation}
\label{eq:bzeta1}
	b_{\zeta1}=\frac{\zeta_1p_0^2}{2}(\beta^{-1}-1)\,,
\end{equation}
\begin{equation}
    b_{\zeta2}=\frac{\zeta_1p_0^2}{2}(\beta^{-1}+1)\,, 
\end{equation}
\begin{equation}
    b_1={B_1}+\frac{4\Lambda_\parallel^2}{3\bar{r}}(3s+4t)\zeta_1\,, 
\end{equation}
\begin{equation}
    b_2={B_2}+\frac{\zeta_1}{6}\left[{3p_0^2}(\beta^{-1}+1)+\frac{8}{\bar{r}}(3s-2t)\Lambda_\perp^2\right]\,, 
\end{equation}
\begin{equation}
    b_3={B_3+B_4}+\frac{\zeta_1\Lambda_\perp^2}{3\bar{r}}(4t-6s) \,, 
\end{equation}
\begin{equation}
    b_4={B_4}\,, 
\end{equation}
\begin{equation}
    \label{eq:b5}
    b_5={B_2}+\frac{\zeta_1}{6}\left[{3p_0^2}(\beta^{-1}-1)-\frac{4}{\bar{r}}(3s+4t)\Lambda_\parallel^2\right]\,.
\end{equation}
Note that the ratio $b_{\zeta 2}/b_{\zeta 1}=\mathcal{R}$ is a consequence of the rotation invariance of the elastomer. Further, as discussed in the earlier sections, these terms would be absent in passive elastomers. We also see that $b_2$ and $b_5$ differ only due to activity (as was previously noted in~\cite{AM_sol}). Their equality in the zero-activity limit is a consequence of Maxwell-Betti reciprocity obeyed by equilibrium solids. As discussed in Sec.~\ref{subsec:achioddel} this does not hold in active systems due to an achiral odd elasticity allowed in uniaxial elastomers.

Almost all the coefficients listed in~\eqref{eq:bzeta1}-\eqref{eq:b5} can change sign because of activity. This can destabilise the uniaxial elastomer. However, such instabilities are system-specific; in this article, our primary focus is on \emph{generic} instabilities driven by chirality. Therefore, we will not discuss the instabilities that arise when $b_{\zeta_1}-b_5$ turn negative. 

Using the fact that $\dot{{\bf u}}={\bf v}$, and decomposing the displacement field perpendicular to the $\hat{{\bf z}}$ axis in components along the in-plane wavevector ${\bf q}_\perp$
\begin{equation}
u_l=\frac{q_xu_x+q_yu_y}{|{\bf q}_\perp|}\,,
\label{eq:ul}
\end{equation}
and transverse to it
\begin{equation}
    u_t=\frac{q_xu_y-q_yu_x}{|{\bf q}_\perp|}\,,
    \label{eq:ut}
\end{equation}
we write~\eqref{zvel1} and~\eqref{perpvel1} explicitly as the dynamics of ${{\bf u}}$, in Fourier space and in the simplified single even viscosity limit obtained by taking $\eta_1=\eta_4=\eta_5=\eta$ and $\eta_2=\eta_3=0$,
\begin{widetext}
\begin{multline}
    \partial_tu_l= -\frac{q_z^2}{\Delta q^4}\left\{b_\zeta^cq_\perp^2\nu_c+\eta[b_3+b_4-b_2)q_\perp^2+b_{\zeta_2}q_z^2\right\}u_l +\frac{q_z^2}{\Delta q^4}\left[(b_4q_\perp^2+b_{\zeta_2}q_z^2)\nu_c-b^c_2 q_\perp^2\eta\right] u_t \\+\frac{q_z|q_\perp|}{\Delta q^4}[b_{\zeta_1}q_\perp^2+(b_1-b_5)q_z^2]\eta u_z\,,
    \label{ulelas}
\end{multline}
\begin{multline}
    \partial_tu_t= -\frac{1}{\Delta q^4}\{q_z^2[(b_3+b_4-b_2)q_\perp^2+b_{\zeta_2}q_z^2]\nu_c-q^2b^c_2 q_\perp^2\eta\}u_l -\frac{1}{\Delta q^4}[\nu_cb^c_2q_z^2q_\perp^2+q^2(b_4q_\perp^2+b_{\zeta_2}q_z^2)\eta]u_t\\ +\frac{|q_\perp|q_z}{\Delta q^4}\{[b_{\zeta_1}q_\perp^2+(b_1-b_5)q_z^2]\nu_c\}u_z\,,
    \label{utelas}
\end{multline}
\begin{multline}
    \partial_tu_z= \frac{q_z|q_\perp|}{\Delta q^4}\left\{b^c_2 q_\perp^2\nu_c+[b_{\zeta_1}q_z^2+(b_3+b4-b_2)q_\perp^2]\eta\right\}u_l - \frac{q_z|q_\perp|}{\Delta q^4}[(b_4q_\perp^2+b_{\zeta_2}q_z^2\nu_c-b^c_2 q_\perp^2\eta]u_t \\-\frac{q_\perp^2}{\Delta q^4}[b_{\zeta_1}q_\perp^2+(b_1-b_5)q_z^2]\eta u_z\,,
    \label{uzelas}
\end{multline}
\end{widetext}
where
\begin{equation}
    \Delta=\frac{\nu_c^2q_z^2}{q^2}+\eta^2\,,
\end{equation}
and 
\begin{equation}
    \nu_c=\frac{\eta^c_2q_\perp^2+\eta^c_1(q_z^2-q_\perp^2)}{q^2}\,. 
\end{equation}
                
Examining~\eqref{ulelas}-\eqref{uzelas}, we see that because of the overall incompressibility, one of the three eigenfrequencies implied by this dynamics vanishes. The other two have both oscillatory and relaxation parts, in general, even at $\mathcal{O}(q^0)$, as in polar and chiral columnar materials~\cite{SJ_col, AM_chi_rev}. Importantly, the oscillatory part vanishes for a wavevector with $q_z=0$. Further, as in chiral columnar materials~\cite{SJ_col}, the oscillations for wavevectors purely along $\hat{{\bf z}}$, $\omega_\pm=\pm b_{\zeta_2}/(\eta^c_1\mp i\eta)$, are solely due to odd velocity-dependent force densities and not due to odd elasticity---a consequence of there being only one odd elastic term. As has been noted~\cite{Xing_annals}, $u_z$ fluctuations in passive uniaxial elastomers are like displacement fluctuations in smectics, and $u_l$ and $u_t$ fluctuations are analogous to displacement fluctuations in columnar materials. It is amusing to find that this continues to hold even upon the inclusion of chirality and activity: there is no chiral force density at this order in gradients in momentum-conserved, active smectic liquid crystals~\cite{SJ_smec} and there is only one such force density---which has the same form as here---in active columnar materials~\cite{SJ_col}.

\subsection{Inertial equations for momentum-conserved active chiral elastomers}
\label{elasin}
Here, for completeness, we display the dynamical equations for active and chiral polar elastomers in the inertial limit in which viscosity can be ignored. These equations of motion can be used for calculating the analogues of the second sound for chiral and polar active elastomers. Plugging in the elastic force densities that appear on the right-hand side of~\eqref{zvel1} and~\eqref{perpvel1} in~\eqref{inertial}, we obtain 
\begin{equation}
    \rho_0 \partial^2_t u_z = b_{\zeta1} \nabla_\perp^2{u}_z + b_1 \partial_z^2 u_z + b_2 \partial_z \nabla_\perp \cdot {\bf u}_\perp \,,
\end{equation}
\begin{multline}
    \rho_0\partial^2_t{\bf u}_\perp= b_{\zeta2}\partial_z^2{\bf u}_\perp+b_3\nabla_\perp\nabla_\perp\cdot{\bf u}_\perp+b_4\nabla_\perp^2{\bf u}_\perp\\+b_5\nabla_\perp\partial_zu_z+b^c_2\nabla_\perp^2\bm{\varepsilon}\cdot{\bf u}_\perp\,,
\end{multline}
where $\rho_0$ is the mean density; the fluctuations of the density field in this material that has no vacancies is completely determined by dilations and contractions of the material: $\delta\rho=-\rho_0\nabla\cdot{\bf u}$. These equations can be written in terms of $u_l$, $u_t$---defined in Sec.~\ref{elasfl}---and $u_z$ as
\begin{gather}
    \rho_0\partial_t^2 u_l = -\bigl[ (b_3+b_4) q_\perp^2 + b_{\zeta_2} q_z^2 \bigr] u_l - b^c_2 q_\perp^2 u_t - b_5|q_\perp|q_z  u_z\,, \\
    \rho_0\partial_t^2 u_t = b^c_2 q_\perp^2 u_l -\bigl( b_4 q_\perp^2 + b_{\zeta_2} q_z^2 \bigr) u_t \,, \\
    \rho_0\partial_t^2 u_z = -b_2 |q_\perp| q_z u_l - \bigl( b_{\zeta_1} q_\perp^2 + b_1 q_z^2 \bigr) u_z\,.
\end{gather}
We now calculate the eigenfrequencies implied by these equations for fluctuations with wavevectors either along $\hat{{\bf z}}$ or orthogonal to it, i.e., for $|q_\perp|=0$ or $q_z=0$. For $q_z=0$, the three pairs of eigenfrequencies are
\begin{equation}
    {\omega_1}_\pm=\pm |q_\perp|\sqrt{{b}_{\zeta_1}/\rho_0}\,,
\end{equation}
\begin{equation}
    {\omega_2}_\pm=\pm \frac{|q_\perp|}{\sqrt{2\rho_0}}\sqrt{{b}_3+2{b}_4+\sqrt{{b}_3^2-(2{b}^c_2)^2}}\,,
\end{equation}
\begin{equation}
    {\omega_3}_\pm=\pm \frac{|q_\perp|}{\sqrt{2\rho_0}}\sqrt{{b}_3+2{b}_4-\sqrt{{b}_3^2-(2{b}^c_2)^2}}\,.
\end{equation}
These eigenfrequencies describe sound waves when ${b}_3>|2{b}^c_2|$ and imply an odd-elasticity-induced instability of the solid state otherwise. The instability in inertial systems is caused by the same force density that led to oscillations in Sec.~\ref{elasfl} and has been discussed in the context of two-dimensional odd solids~\cite{Deboo_viscel, AM_chi_rev}. Here, it appears in a three-dimensional solid state. For $|q_\perp|=0$, the three sets of eigenfrequencies are 
\begin{equation}
\label{eigen_iner}
    {\omega_1}_\pm=\pm q_z\sqrt{\frac{{b}_{1}}{\rho_0}}\,, \quad {\omega_2}_\pm={\omega_3}_\pm=\pm q_z\sqrt{\frac{{b}_{\zeta_2}}{\rho_0}}\,.
\end{equation}
We thus find that, as expected, there is no effect of chirality for perturbations with wavevectors along $\hat{{\bf z}}$. These modes describe sound waves (except in the case in which some of the elastic coefficients, for instance, ${b}_{\zeta_2}$ turn negative). Interestingly, two pairs of modes have the same dispersion relation. We will see that chirality lifts this degeneracy in uniaxial active \emph{solids} in Sec.~\ref{subsec:inertial_solids}, that is, the degeneracy is a result of the additional symmetry in elastomers.

\section{From elastomers to solids via a symmetry breaking field}
\label{sec:Unisol}
As we have mentioned throughout this article, a uniaxial elastomer is an idealised high-symmetry phase that is invariant under \emph{independent} rotations of the ordering direction and the solid structure, i.e., has $SO(3)\times SO(3)$ symmetry. General uniaxial solids are less symmetric (i.e., in a sense, they are \emph{more} ordered): they are only invariant under the \emph{joint} rotation of the ordering direction and the solid structure, i.e., they only have $SO(3)$ symmetry. Further, it is difficult to realise the uniaxial elastomeric phase: elastomers that are crosslinked in the nematic state~\cite{Ranjan_Lubensky} retain the memory of the state that they are crosslinked in, and this breaks the higher $SO(3)\times SO(3)$ symmetry to $SO(3)$. In this section, we will examine the effect of chiral activity on uniaxial (both polar and nematic) solids by constructing the linearised hydrodynamic equations of motion of this phase. Crucially---and distinctly from the accompanying Letter~\cite{short} where these equations were constructed directly based on symmetry considerations---we will do so by starting with our theory for chiral and active uniaxial \emph{elastomers} and explicitly introducing a field that breaks $SO(3)\times SO(3)$ symmetry to $SO(3)$. Physically, this field may represent the effect of cross-linking in a state with a mean direction ${\bf n}_0$.

As in uniaxial elastomers, orientational fluctuations are massive in uniaxial solids and therefore are not hydrodynamic degrees of freedom. However, the manner in which they are eliminated in favour of displacement fluctuations is distinct because of the distinct operations the two states are invariant under. The distinct relations of the director fluctuations to displacement fluctuations are known to lead to different hydrodynamic mode structures for passive solids and elastomers~\cite{Ranjan_Lubensky, Stenull_lub_hydro} and, as we shall show, for chiral and active uniaxial materials as well.

The rest of this section is organised as follows: in Sec.~\ref{subsec:Slavsol} we explicitly introduce a field that breaks the higher elastomeric symmetry down to that of a uniaxial solid. We show how the orientational fluctuations are determined in terms of displacement fluctuations in the presence of this field. In Sec.~\ref{dynsolfric}, we construct the dynamical equations for a chiral and polar active solid moving through a fixed background medium. This corresponds to the case discussed in detail in~\cite{short}. We show that chiral activity generically destabilises such a solid. In Sec.~\ref{dyneqsolfl}, we discuss the dynamics of a polar and chiral solid in a Stokesian fluid medium. In Sec.~\ref{subsec:inertial_solids}, we discuss the non-dissipative dynamics of such solids. Finally, in Sec.~\ref{subsec:uniaxial_elasticity}, we summarise a group-representation theory-based description of the elasticity of uniaxial passive and active solids.

\subsection{Elimination of orientational fluctuations and discussion of free energy in uniaxial solids}
\label{subsec:Slavsol}
We now show how orientational fluctuations relax to a value determined by the local displacement field gradients in uniaxial solids. For this, we start with the free energy in~\eqref{FU}, where the displacement field is already defined relative to a uniaxial state, and add an additional symmetry-breaking field $\propto h$ that ensures that the free energy remains invariant under the \emph{joint} rotation of the director and the elastic material but not under their independent rotations:
\begin{widetext}
    \begin{multline}
		F=\frac{1}{2}\int d^3{\bf x}\Bigg\{B_1\left(E^S_{zz}\right)^2+2B_2E^S_{zz}\left(E^S_{xx}+E^S_{yy}\right)+B_3\left(E^S_{xx}+E^S_{yy}\right)^2+2B_4\left({E^S_{xx}}^2+{E^S_{yy}}^2+2{E^S_{xy}}^2\right)\\+\bar{\mu}\left[\beta\left(\delta{\bf p}_\perp/p_0-\bm{E}^A_{z\perp}\right)-\bm{E}^S_{z\perp}\right]^2+h \left(\delta{\bf p}_\perp/p_0-\bm{E}^A_{z\perp}\right)^2\Bigg\}\,.
        \label{Fsymbrkn}
	\end{multline}
\end{widetext}
This implies that for $h\neq 0$
\begin{equation}
    \delta{\bf p}_\perp = p_0 \left( \bm{E}^A_{z\perp} + \frac{\beta\bar{\mu}}{h+\beta^2\bar{\mu}} \bm{E}^S_{z\perp} \right)\,,
\label{dircross}
\end{equation}
which reduces to~\eqref{pperpu} in the $h\to 0$ limit. To leading order in $h$, this reads
\begin{equation}
    \delta{\bf p}_\perp = p_0 \left( \bm{E}^A_{z\perp} + \beta^{-1}\bm{E}^S_{z\perp} - \frac{h}{\beta^3\bar{\mu}} \bm{E}^S_{z\perp} \right)\,.
\label{dircross2}
\end{equation}

Using~\eqref{dircross} in~\eqref{Fsymbrkn}, we see that now there is a cost for $\bm{E}^S_{z\perp}$ deformations even at the harmonic order:
\begin{widetext}
\begin{equation}
    \begin{split}
	   F=\frac{1}{2}\int d^3{\bf x} \biggl\{ & B_1\left(E^S_{zz}\right)^2 + 2B_2E^S_{zz}\left(E^S_{xx}+E^S_{yy}\right)+B_3\left(E^S_{xx}+E^S_{yy}\right)^2 \\
       & \quad +2B_4\left({E^S_{xx}}^2+{E^S_{yy}}^2+2{E^S_{xy}}^2\right)+2B_5\left({E^S_{xz}}^2+{E^S_{yz}}^2\right)\biggr\}\,,
    \end{split}
    \label{eq:Fsol}
\end{equation}
\end{widetext}
with 
\begin{equation}
    B_5=\frac{h\bar{\mu}}{2(h+\beta^2\bar{\mu})} \,,
    \label{eq:B5h}
\end{equation}
vanishing for $h\to 0$. 

As earlier, the passive, linearised elastic force densities for uniaxial solids are given by $-\delta F/\delta{\bf u}$, where $F$ is the energy in~\eqref{eq:Fsol}. These force densities can also be written as divergences of stresses: $-\delta F/\delta{\bf u}=\nabla\cdot\bm{\sigma}^e$.

Another important consequence of~\eqref{dircross} and~\eqref{dircross2} is that $\delta{\bsf Q}_{z\perp}$ is no longer $\propto\delta{\bsf C}_{z\perp}$ where ${\bsf C}$ is the strain tensor is about a fiducial isotropic elastic material. Using~\eqref{Wzpu} and~\eqref{lamaniso} we can see from~\eqref{dircross2} that 
\begin{equation}
    \delta{\bsf Q}_{z\perp}=\frac{\mu}{t}\delta{\bsf C}_{z\perp}-\frac{hp_0^2}{\beta^3\bar{\mu}}\bm{E}^S_{z\perp}\,.
\end{equation}
From~\eqref{Wzpu}, we see that $\bm{E}^S_{z\perp}$ is not simply proportional to ${\bsf U}_{z\perp}$ and hence not directly proportional to ${\bsf C}_{z\perp}$. Our conclusion that the chiral force density in~\eqref{frc} or the odd elasticity with coefficient $\zeta^c_1$ in~\eqref{cstrs} does not contribute to linear order in a polar elastomeric state required $\delta{\bsf Q}_{z\perp}$ to be proportional to $\delta{\bsf C}_{z\perp}$. Since this is not the case in a uniaxial solid, we expect that the force densities with the coefficient $w^c_1$ and $\zeta^c_1$ contribute to linear order in a uniaxial solid state. We show that this expectation is borne out by constructing the equations of motion for uniaxial solids in various dynamical regimes described in Sec.~\ref{Dynregime} and explicitly demonstrate that it is a consequence of $h$. That is, we show how the field $h$ modifies the dynamical equations discussed in Secs.~\ref{elfixback}, \ref{elasfl} and~\ref{elasin} to leading order in $h$.

\subsection{Dynamical equations in a fixed, external background medium}
\label{dynsolfric}
We now turn to the dynamical equations in the polar solid phase in chiral active systems. First, we examine how the dynamical equations are modified from Sec.~\ref{elfixback} in the presence of $h$ when the solid self-propels through a fixed background medium with a speed $v_0$ that remains unchanged by $h$ and is given by~\eqref{speed}. In a frame moving with the mean velocity of the solid, the displacement field equations read
\begin{gather}
    \partial_t u_x=\bar{a}_3\partial_z u_x+\bar{a}_4\partial_x u_z+\bar{a}^c(\partial_z u_y+\partial_yu_z)+\mathcal{O}(\nabla^2)\,,
    \label{uxad} \\
    \partial_t u_y=\bar{a}_3\partial_z u_y+\bar{a}_4\partial_yu_z-\bar{a}^c(\partial_zu_x+\partial_xu_z)+\mathcal{O}(\nabla^2)\,,
    \label{uyad} \\
    \partial_t u_z=\bar{a}_1\partial_z u_z+\bar{a}_2\nabla_\perp\cdot{\bf u}_\perp+\mathcal{O}(\nabla^2)\,.
    \label{uzad}
\end{gather}
These equations are equivalent to the ones discussed in~\cite{short}. The coefficients $\bar{a}_1$, $\bar{a}_2$, $\bar{a}_3$ and $\bar{a}_4$ reduce to $a_1$, $a_2$, $a_3$ and $a_4$ in Eqs.~\eqref{eq:a1}, \eqref{eq:a2}, \eqref{eq:a3} and~\eqref{eq:a4} respectively to leading order in $h$ as these are non-zero when $h=0$. Using~\eqref{dircross} in~\eqref{frc} we find that $\bar{a}^c$ \emph{does not} vanish when $h\neq0$:
\begin{equation}
    \bar{a}^c=\frac{hp_0^2w^c_1(\Lambda_\parallel^2+\Lambda_\perp^2)}{4(h+\beta^2\bar{\mu})}\approx\frac{hp_0^2w^c_1(\Lambda_\parallel^2+\Lambda_\perp^2)}{4\beta^2\bar{\mu}}\,,
\end{equation}
where the final expression is the value of $\bar{a}^c$ to the leading order in $h$.
    
\begin{figure}[t]
    \centering
    \includegraphics[width=\linewidth]{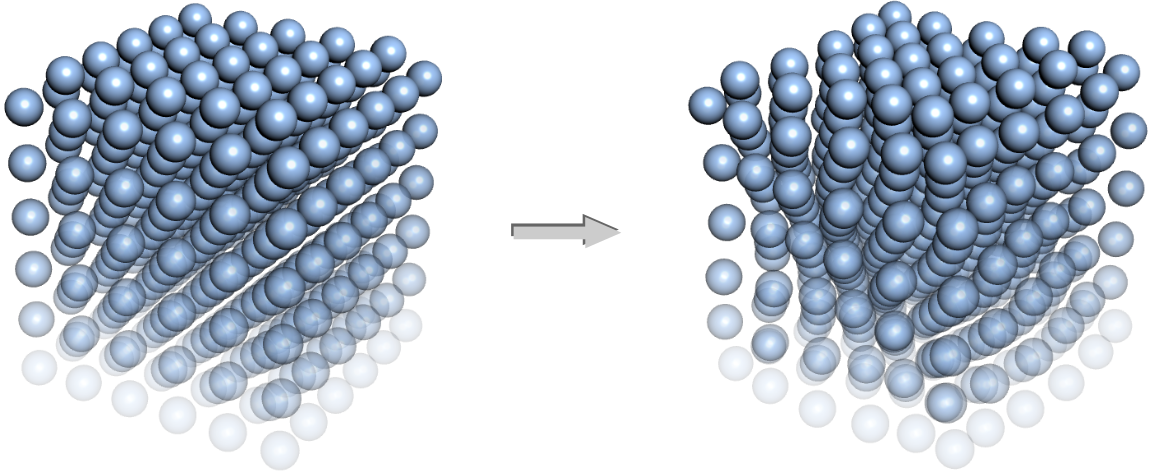}
    \caption{Linear instability of chiral active solids. The solid is shown as a lattice of spheres with the unperturbed state on the left and the linearly unstable displacement mode associated to~\eqref{eq:unstable_solid_frequency} on the right.}
    \label{fig:lattice}
\end{figure}

As discussed in~\cite{short}, the chiral solid phase is generically destabilised by $\bar{a}^c$: considering a perturbation purely along $\hat{{\bf z}}$, we see that $u_z$ dynamics decouples from $u_x$ and $u_y$ and has an eigenfrequency $\omega_z=-\bar{a}_1q_z$. The coupled $u_x$ and $u_y$ dynamics have the eigenfrequencies 
\begin{equation}
    \omega_\pm=-q_z(\bar{a}_3\pm i\bar{a}^c)\,.
    \label{eq:unstable_solid_frequency}
\end{equation}
Since the imaginary part of this eigenfrequency is positive for positive (negative) $q_z$ when $\bar{a}^c<0$ ($\bar{a}^c>0$) and since there is no other imaginary contribution at this or lower order in wavenumbers, the solid is \emph{generically} unstable, without a threshold in $\bar{a}^c$ (i.e., for arbitrarily small values of the symmetry-breaking term $h$) with a linear growth rate. The instability is cut off at larger wavenumbers by elasticity, which contributes at $\mathcal{O}(q^2)$. As we discussed, all physical realisations of uniaxial elastomers have a non-zero $h$ \cite{Ranjan_Lubensky}; therefore, the interplay of activity and chirality generically destabilises such materials at lengthscales $\sim 1/h$. The linear displacement mode of the instability is illustrated in Fig.~\ref{fig:lattice}. Interestingly, while more popular examples of generic instability of ordered states in active materials~\cite{Aditi, SJ_smec, RMP, AM_interface, Rayan} arise in systems with long-range interactions due to Stokesian hydrodynamics, the helical instability we discuss here arises in the opposite dry limit.

We now compare this generic instability with certain other superficially similar chiral instabilities. In Ref.~\cite{RJoy}, a theory of chiral active filaments was constructed. Since the strain field of the three-dimensional solid that is essential to our consideration is absent in the filament, there is no analogue of the generic instability we uncover in that case. The authors of that article, however, show that the combination of chirality and activity leads to a helical contribution to the shape fluctuation spectrum at $\mathcal{O}(q^3)$ that tends to be destabilising. This can be understood as follows: consider a filament in the Monge gauge with fluctuations about a fiducial $z$ axis being denoted by ${\bf h}\equiv(h_x,h_y)$. The lowest-order-in-gradient contribution of chiral activity to the dynamics of ${\bf h}$ is given by $\partial_t{\bf h}\propto\partial_z^3\bm{\varepsilon}\cdot{\bf h}$. This contribution is destabilising. However, it appears at  $\mathcal{O}(q_z^3)$, unlike the instability in this work whose growth rate is $\propto q$; $\partial_z{\bf h}$ corresponds to a uniform tilt of the filament and therefore, a term $\propto\partial_z\bm{\varepsilon}\cdot{\bf h}$ which, if it existed, would have resulted in an instability with a growth rate linear in wavenumbers, is forbidden by rotation invariance. Because the destabilising effect of chirality appears at higher order in wavenumbers for shape fluctuations of filaments, it can potentially be suppressed by an actively generated line tension of the filament~\cite{Norio}, unlike the instability we discuss in this work, which cannot be suppressed by elasticity at small wavenumbers. An active and chiral filament \emph{does} suffer an instability with a growth rate linear in wavenumber when it is not invariant under rotation, for example, due to being put in a frame. This is curious since, for passive filaments, being put in a frame reduces fluctuations because of the generation of a line tension. 

A comparison of the chiral activity-induced generic instability of the unaxial solid that we uncover here with the instability of the active or passive uniaxial state (either polar or nematic) in chiral systems discussed in Sec.~\ref{instab_unifl} is also useful. Indeed, the two instabilities look superficially similar. To see this explicitly, consider the minimal equation for the polarisation field in three-dimensional, dry, chiral active systems that we discussed in~\eqref{eq:polchi}. The equation of motion for fluctuations about a polar state~\eqref{eq:flucpolchi} has the same form as~\eqref{uxad} and~\eqref{uyad} when only fluctuations along $\hat{{\bf z}}$ are considered. Note further that the chiral terms are invariant under ${\bf p}\to -{\bf p}$. However, the physics encoded in~\eqref{uxad} and~\eqref{uyad} is very different from the one described by~\eqref{eq:flucpolchi}. The former describes the dynamics of the displacement field and not the director. Indeed, as discussed in Sec.~\ref{sec:passens}, the director field of a uniaxial \emph{solid} does not suffer an instability in chiral systems in three dimensions; instead, it relaxes to a value determined by the gradients of the displacement field  with a rate that is non-vanishing in the limit of zero wavenumbers. 

When chirality induces a helical instability of the director field of a uniaxial \emph{fluid} the state that the system finally settles into has a helical director configuration~\cite{deGen}. Though chirality-induced helical instability of the displacement fields of the uniaxial solid phase is distinct, we speculate that the system may settle into a state with an orientation field like that of a cholesteric.

\subsection{Dynamical equations in a fluid medium}
\label{dyneqsolfl}
In this section, we display the dynamical equations for a polar and chiral {solid} in a Stokesian fluid, to the leading order in gradients. In particular, we show that the presence of $h$, which breaks the elastomeric symmetry, allows for an additional odd elastic force density. The starting dynamical equation is the same as~\eqref{fbal}, with $\nabla\cdot{\bf v}=0$ and $F$ now being given by~\eqref{eq:Fsol}.

The force balance equation reads 
\begin{widetext}
    \begin{equation}
    \label{solfulleqz}
        0 = -\partial_z \Pi + \bar{b}_{\zeta_1} \nabla_\perp^2 {u}_z + \eta_5 \nabla_\perp^2 {v}_z + (\bar{b}_1 \partial_z^2 u_z + \eta_1 \partial_z^2 v_z) + \bar{b}_2 \partial_z \nabla_\perp \cdot {\bf u}_\perp + \eta_2 \partial_z \nabla_\perp \cdot {\bf v}_\perp + \bar{b}^c_1 \partial_z \nabla_\perp \times {\bf u}_\perp + \eta^c_1 \partial_z \nabla_\perp \times {\bf v}_\perp\,,
    \end{equation}
    \begin{multline}
    \label{solfulleqperp}
        0 = -\nabla_\perp \Pi + \bar{b}_{\zeta_2} \partial_z^2 {\bf u}_\perp + \eta_5 \partial_z^2 {\bf v}_\perp + \bar{b}_3 \nabla_\perp \nabla_\perp \cdot {\bf u}_\perp + \eta_3 \nabla_\perp \nabla_\perp \cdot{\bf v}_\perp + \bar{b}_4 \nabla_\perp^2 {\bf u}_\perp + \eta_4 \nabla_\perp^2 {\bf v}_\perp + \bar{b}_5 \nabla_\perp \partial_z v_z \\ + \eta_2 \nabla_\perp \partial_z v_z + \bar{b}^c_2 \nabla_\perp^2 \bm{\varepsilon} \cdot {\bf u}_\perp + \eta^c_2 \nabla_\perp^2 \bm{\varepsilon} \cdot {\bf v}_\perp + \bar{b}^c_1 \partial_z \bm{\varepsilon} \cdot (\partial_z {\bf u}_\perp + \nabla_\perp {u}_z) + \eta^c_1 \partial_z \bm{\varepsilon} \cdot (\partial_z {\bf v}_\perp + \nabla_\perp {v}_z)\,.
    \end{multline}
\end{widetext}
Except for $\bar{b}^c_1$, all the barred parameters reduce to the unbarred parameters in~\eqref{eq:bc}-\eqref{eq:b5} to leading (zeroth) order in $h$. Importantly, when $h\neq 0$, $b_{\zeta_1}$ and $b_{\zeta_2}$ are no longer purely active. Instead, they have a passive part $B_5/2$, where the expression of $B_5$ in terms of $h$ is displayed in~\eqref{eq:B5h}. $\bar{b}_2$ and $\bar{b}_5$ also contains an additional passive part $B_5/2$. The coefficient $\bar{b}^c_1$ vanishes in the $h\to 0$ limit. For $h\neq 0$,
\begin{equation}
    \bar{b}^c_1=\frac{hp_0^3\zeta^c_1(\Lambda_\parallel^2+\Lambda_\perp^2)}{2(h+\beta^2\bar{\mu})}\approx\frac{hp_0^3\zeta^c_1(\Lambda_\parallel^2+\Lambda_\perp^2)}{2\beta^2\bar{\mu}}\,,
\end{equation}
where the second approximate equality is for small $h$.

Eliminating the pressure, using $\dot{{\bf u}}={\bf v}$ and writing the equation for displacement field in terms of $u_l$ and $u_t$ defined by~\eqref{eq:ul} and~\eqref{eq:ut}, we get in the single even viscosity limit ($\eta_1=\eta_4=\eta_5=\eta$ and $\eta_2=\eta_3=0$),
\begin{widetext}
    \begin{multline}
        \partial_tu_l= -\frac{q_z^2}{\Delta q^4}\left\{(\bar{b}_1^cq_\perp^2+\bar{b}^c_1q_z^2)\nu_c+\eta[\bar{b}_3+\bar{b}_4-b_2)q_\perp^2+\bar{b}_{\zeta_2}q_z^2]\right\}u_l +\frac{q_z^2}{\Delta q^4}\left[(\bar{b}_4q_\perp^2+\bar{b}_{\zeta_2}q_z^2)\nu_c-q^2\mu_c\eta\right] u_t \\+\frac{q_z|q_\perp|}{\Delta q^4}\left\{-\bar{b}^c_1\nu_cq_z^2+[\bar{b}_{\zeta_1}q_\perp^2+(\bar{b}_1-\bar{b}_5)q_z^2]\eta\right\}u_z\,,
        \label{ulsol}
    \end{multline}
    \begin{multline}
        \partial_tu_t= -\frac{1}{\Delta q^4}\{q_z^2[(\bar{b}_3+\bar{b}_4-\bar{b}_2)q_\perp^2+\bar{b}_{\zeta_2}q_z^2]\nu_c-q^2(\bar{b}^c_2q_\perp^2+\bar{b}^c_1q_z^2)\eta\}u_l -\frac{1}{\Delta q^2}[\nu_c\mu_cq_z^2+(\bar{b}_4q_\perp^2+b_{\zeta_2}q_z^2)\eta]u_t\\ +\frac{|q_\perp|q_z}{\Delta q^4}\{[\bar{b}_{\zeta_1}q_\perp^2+(\bar{b}_1-\bar{b}_5)q_z^2]\nu_c+\bar{b}^c_1q^2\eta\}u_z\,,
        \label{utsol}
    \end{multline}
    \begin{multline}
        \partial_tu_z= \frac{q_z|q_\perp|}{\Delta q^4}\left\{(\bar{b}^c_2q_\perp^2+\bar{b}^c_1q_z^2)\nu_c+[\bar{b}_{\zeta_1}q_z^2+(\bar{b}_3+\bar{b}4-\bar{b}_2)q_\perp^2]\eta\right\}u_l - \frac{q_z|q_\perp|}{\Delta q^4}[(\bar{b}_4q_\perp^2+\bar{b}_{\zeta_2}q_z^2)\nu_c-q^2\mu_c\eta]u_t \\-\frac{q_\perp^2}{\Delta q^4}\{-\bar{b}^c_1q_z^2\nu_c+[\bar{b}_{\zeta_1}q_\perp^2+(\bar{b}_1-\bar{b}_5)q_z^2]\eta\}u_z\,,
        \label{uzsol}
    \end{multline}
\end{widetext}
where
\begin{equation}
    \mu_c=\frac{\bar{b}^c_2q_\perp^2+\bar{b}^c_1(q_z^2-q_\perp^2)}{q^2}\,.
\end{equation}
As advertised, these equations feature two chiral odd elastic terms and not one; these are the terms with coefficients $\bar{b}^c_1$ and $\bar{b}^c_2$ that derive from the active and chiral stresses of Sec.~\ref{subsec:active_elastic_stresses}. One of the three eigenfrequencies vanishes to $\mathcal{O}(q^0)$ due to the constraint of incompressibility, while the other two may have oscillatory parts. The distinction between this and the elastomeric case discussed in Sec.~\ref{elasfl} is most clearly seen for wavevectors along $\hat{{\bf z}}$: 
\begin{equation}
\omega_\pm=-\frac{i(\bar{b}_{\zeta_2}\eta+\bar{b}^c_1\eta^c_1)\pm  (\bar{b}^c_1\eta-\eta^c_1\bar{b}_{\zeta_2})}{\eta^2+{\eta^c_1}^2}\,,
\end{equation}
where the oscillatory part now does not vanish even when $\eta^c_1=0$ i.e., there is a contribution to the oscillations due to odd elasticity even for wavevectors purely along $\hat{{\bf z}}$.

\subsection{Inertial equations for momentum-conserved active chiral solids}
\label{subsec:inertial_solids}
In this section, we display the dynamical equations for active and chiral polar solids in the inertial limit in which viscosities and odd viscosities can be ignored. We further do not assume that the system is incompressible. Using elastic force densities on the right-hand side of~\eqref{solfulleqz} and~\eqref{solfulleqperp} in~\eqref{inertial}, we obtain
\begin{equation}
    \rho_0 \partial^2_t u_z = \bar{b}_{\zeta1} \nabla_\perp^2 {u}_z + \bar{b}_1 \partial_z^2 u_z + \bar{b}_2 \partial_z \nabla_\perp \cdot {\bf u}_\perp + \bar{b}^c_1 \partial_z \nabla_\perp \times {\bf u}_\perp \,,
\end{equation}
\begin{multline}
    \rho_0\partial^2_t{\bf u}_\perp= \bar{b}_{\zeta2}\partial_z^2{\bf u}_\perp+\bar{b}_3\nabla_\perp\nabla_\perp\cdot{\bf u}_\perp+\bar{b}_4\nabla_\perp^2{\bf u}_\perp+\bar{b}_5\nabla_\perp\partial_zu_z\\+\bar{b}^c_2\nabla_\perp^2\bm{\varepsilon}\cdot{\bf u}_\perp+\bar{b}^c_1\partial_z\bm{\varepsilon}\cdot(\partial_z{\bf u}_\perp+\nabla_\perp{u}_z)\,,
\end{multline}
where $\rho_0$ is the mean density. In terms of the displacement field $u_l$, $u_t$ and $u_z$, these equations read
\begin{multline}
    \rho_0\partial_t^2 u_l= -[(\bar{b}_3+\bar{b}_4)q_\perp^2+\bar{b}_{\zeta_2}q_z^2]u_l \\-(\bar{b}^c_2q_\perp^2+\bar{b}^c_1q_z^2) u_t - \bar{b}_5|q_\perp|q_z  u_z\,,
\end{multline}
\begin{equation}
    \rho_0\partial_t^2 u_t= (\bar{b}^c_2q_\perp^2+\bar{b}^c_1q_z^2)u_l -(\bar{b}_4q_\perp^2+\bar{b}_{\zeta_2}q_z^2) u_t + \bar{b}^c_1|q_\perp|q_z  u_z\,,
\end{equation}
\begin{equation}
    \rho_0\partial_t^2 u_z=-\bar{b}_2|q_\perp|q_z u_l - \bar{b}^c_1|q_\perp|q_z u_t -(\bar{b}_{\zeta_1}q_\perp^2+\bar{b}_1q_z^2)   u_z\,.
\end{equation}
The dispersion relation for a general direction is quite involved. Therefore, we display the dispersion relations for wavevectors either along $\hat{{\bf z}}$ or orthogonal to it, i.e., for $|q_\perp|=0$ or $q_z=0$. For $q_z=0$, the three sets of eigenfrequencies are
\begin{gather}
    {\omega_1}_\pm=\pm |q_\perp|\sqrt{\frac{\bar{b}_{\zeta_1}}{\rho_0}}\,, \\
    {\omega_2}_\pm=\pm \frac{|q_\perp|}{\sqrt{2\rho_0}}\sqrt{\bar{b}_3+2\bar{b}_4+\sqrt{\bar{b}_3^2-(2\bar{b}^c_2)^2}}\,, \\
    {\omega_3}_\pm=\pm \frac{|q_\perp|}{\sqrt{2\rho_0}}\sqrt{\bar{b}_3+2\bar{b}_4-\sqrt{\bar{b}_3^2-(2\bar{b}^c_2)^2}}\,.
\end{gather}
These eigenfrequencies describe sound waves when $\bar{b}_3>|2\bar{b}^c_2|$ and imply an odd-elasticity-induced instability of the solid state otherwise (just as in Sec.~\ref{elasin}). For $|q_\perp|=0$, the three sets of eigenfrequencies are 
\begin{align}
    {\omega_1}_\pm & = \pm q_z\sqrt{\frac{\bar{b}_{1}}{\rho_0}}\,, \\
    \label{eigen1}
    {\omega_2}_\pm & = \pm q_z\sqrt{\frac{\bar{b}_{\zeta_2}+i\bar{b}^c_1}{\rho_0}}\,, \\
    \label{eigen2}
    {\omega_3}_\pm & = \pm q_z\sqrt{\frac{\bar{b}_{\zeta_2}-i\bar{b}^c_1}{\rho_0}}\,.
\end{align}
Note that the odd elastic term $\propto \bar{b}^c_1$ which vanishes when $h=0$, lifts the degeneracy of the two eigenfrequencies in~\eqref{eigen_iner}. Moreover, \eqref{eigen1} and~\eqref{eigen2} implies that, because of $\bar{b}^c_1$, two eigenfrequencies have positive imaginary parts that are linear in $q_z$. This implies that momentum-conserved, three-dimensional odd solids are \emph{generically} unstable in the inertial limit with a linear growth rate. Indeed, this instability is caused by the same force density that leads to {oscillations} in the inertia\emph{less} limit. Since $\bar{b}^c_1$ is absent in uniaxial \emph{elastomers} (see Sec.~\ref{elasin}), this is another case in which the higher elastomeric symmetry protects that state from a chirality-induced instability.

\subsection{Elasticity of uniaxial solids using group representations}
\label{subsec:uniaxial_elasticity}

In this final subsection we summarise a group-representation-theoretic description of the elasticity of uniaxial solids. For passive elasticity the results are classical (e.g.~\cite{LandauLifshitz}) but our approach emphasises the structure of the odd elasticity in relation to material symmetries and provides a complement to the rest of this article. Linear elasticity is a relation between strain and stress, the symmetric part of which is controlled by a free energy function and Maxwell-Betti reciprocity~\cite{maxwell1864,betti1872}; antisymmetric contributions are allowed in active systems and represent non-reciprocal odd elasticities~\cite{scheibner2020odd}.  

Uniaxial solids are characterised by a single preferred direction or axis, which we take to be along $\hat{\bf z}$, and about which they are isotropic. Their symmetry group is $C_{\infty}$ if the axis is polar and $D_{\infty}$ if it is apolar. (We exclude any mirror planes as we are considering chiral materials.) Their elasticity is a linear relation between the symmetrised linear strain and the symmetric part of the stress that is constrained only by the uniaxial symmetry. It can be determined from the structure of the spaces of stress and strain that is induced by the action of the symmetry group. 

The linearised strain $U_{ij} = \frac{1}{2} ( \partial_i u_j + \partial_j u_i )$ belongs to a six-dimensional vector space, as does the symmetric part of the stress. The action of the rotation group $SO(3)$ (of Euclidean space) splits this as $6 = 1 \oplus 5$, where the $1$ corresponds to a one-dimensional trivial representation spanned by the identity and the $5$ by the deviatoric strain, giving the decomposition ${\bsf U} = (1/3)\Phi {\bsf I} + {\bsf C}$. For a uniaxial solid the deviatoric part splits further into irreducible representations (irreps) as $5 = 1 \oplus 2_1 \oplus 2_2$, where the subscripts on the two (non-trivial) two-dimensional irreps correspond to their spin. Significantly, the one-dimensional part is again a trivial representation so that symmetry does not distinguish it from the irrep spanned by the identity. 

We introduce an orthonormal basis ${\bsf S}_a$, $a=1, \dots, 6$ adapted to this decomposition into irreps 
\begin{align}
    & {\bsf S}_1 = \frac{1}{\sqrt{3}} \,{\bsf I} , && {\bsf S}_2 = \frac{1}{\sqrt{6}} \bigl( 3 \,\hat{\bf z} \hat{\bf z} - {\bsf I} \bigr) , \\
    & {\bsf S}_3 = \frac{1}{\sqrt{2}} \bigl( \hat{\bf z} \hat{\bf x} + \hat{\bf x} \hat{\bf z} \bigr) , && {\bsf S}_4 = \frac{1}{\sqrt{2}} \bigl( \hat{\bf z} \hat{\bf y} + \hat{\bf y} \hat{\bf z} \bigr) , \\
    & {\bsf S}_5 = \frac{1}{\sqrt{2}} \bigl( \hat{\bf x} \hat{\bf x} - \hat{\bf y} \hat{\bf y} \bigr) , && {\bsf S}_6 = \frac{1}{\sqrt{2}} \bigl( \hat{\bf x} \hat{\bf y} + \hat{\bf y} \hat{\bf x} \bigr) ,
\end{align}
where $\{ {\bsf S}_1, {\bsf S}_2 \}$ span the trivial $1 \oplus 1$ representation, $\{ {\bsf S}_3, {\bsf S}_4 \}$ span the $2_1$ irrep and $\{ {\bsf S}_5, {\bsf S}_6 \}$ span the $2_2$ irrep. The linearised strain can be written in this basis as ${\bsf U} = \sum_a U_a {\bsf S}_a$, where $U_a = \mathrm{Tr} \,{\bsf S}_a {\bsf U}$, and similarly for the symmetric part of the stress. Symmetry does not allow mixing between these three two-dimensional subspaces, so that the stress-strain relation can be given separately for each of them. As the two one-dimensional irreps are both the trivial representation, the symmetry group places no constraint on the form of the linear relation between strain and stress within this two-dimensional subspace and it therefore takes the general form (in the $\{ {\bsf S}_1, {\bsf S}_2 \}$ basis) 
\begin{equation}
    \begin{split}
        & B^{1\oplus 1}_{0} \begin{bmatrix} 1 & 0 \\ 0 & 1 \end{bmatrix} + B^{1\oplus 1}_{+} \begin{bmatrix} 1 & 0 \\ 0 & -1 \end{bmatrix} + B^{1\oplus 1}_{\times} \begin{bmatrix} 0 & 1 \\ 1 & 0 \end{bmatrix} \\
        &  \quad + B^{1\oplus 1}_{\textrm{odd}} \begin{bmatrix} 0 & -1 \\ 1 & 0 \end{bmatrix} .
    \end{split}
    \label{eq:elasticity_1+1}
\end{equation}
There are three symmetric components and one antisymmetric component, associated with the generator of rotations (or complex structure) within this two-dimensional subspace and corresponding to an odd elasticity~\cite{shankar2024active}. Interestingly, as ${\bsf S}_1$ and ${\bsf S}_2$ are both apolar---invariant under $\hat{\bf z} \to - \hat{\bf z}$ (either by reflection or by rotation about a horizontal axis)---this odd modulus is present even in apolar and achiral uniaxial active solids with $D_{\infty}$ (or $D_{\infty h}$) symmetry. This point was also made following~\eqref{noreciachi}, where the invariance under ${\bf p} \to - {\bf p}$ is manifest. The symmetric part of~\eqref{eq:elasticity_1+1} can be diagonalised by a change of basis and expressed equivalently as two moduli $B^{1\oplus 1}_0 \pm \sqrt{(B^{1\oplus 1}_{+})^2 + (B^{1\oplus 1}_{\times})^2}$, both of which should be positive in a passive material, and an angle giving the change of basis relative to $\{ {\bsf S}_1, {\bsf S}_2 \}$~\cite{LandauLifshitz}. We denote the diagonal basis by $\{ {\bsf D}_1 , {\bsf D}_2 \}$ and write $B^{1\oplus 1}_1 = \sqrt{(B^{1\oplus 1}_{+})^2 + (B^{1\oplus 1}_{\times})^2}$. The orientation of the diagonal basis is commonly given relative to the basis $\{ \tfrac{1}{\sqrt{2}} ( \hat{\bf x} \hat{\bf x} + \hat{\bf y} \hat{\bf y} ) , \hat{\bf z} \hat{\bf z} \}$ (rather than $\{ {\bsf S}_1 , {\bsf S}_2 \}$) as a reference, which we adopt, denoting by $\theta$ the angle between these two frames. This change of basis does not affect the identity or complex structure parts of~\eqref{eq:elasticity_1+1}, which are independent of the choice of basis provided the orientation is maintained. 

Within the two two-dimensional irreps the relation between strain and stress is constrained by the non-triviality of the group action and only the identity and complex structure parts of the linear transformation are compatible with the action of the symmetry group. The odd terms $B^{2_1}_{\textrm{odd}}$ and $B^{2_2}_{\textrm{odd}}$ are further only allowed in a polar material, since flipping the uniaxial direction results in changing the sign of both coefficients. These odd moduli correspond to $\zeta^c_1$ and $\zeta^c_2$, respectively, in~\eqref{cstrs}. 

For completeness, we summarise the elasticity of uniaxial solids by giving explicitly the general form of the stress-strain relation (for $C_{\infty}$ symmetry)  
\begin{widetext}
\begin{equation}
\begin{split}
    \boldsymbol{\sigma} & = \bigl( B^{1\oplus 1}_{0} + B^{1\oplus 1}_{1} \bigr) \biggl[ \frac{\cos\theta}{\sqrt{2}} \bigl( \partial_x u_x + \partial_y u_y \bigr) + \sin\theta \,\partial_z u_z \biggr] {\bsf D}_1 + \bigl( B^{1\oplus 1}_{0} - B^{1\oplus 1}_{1} \bigr) \biggl[ \cos\theta \,\partial_z u_z - \frac{\sin\theta}{\sqrt{2}} \bigl( \partial_x u_x + \partial_y u_y \bigr) \biggr] {\bsf D}_2 \\
    & + B^{1\oplus 1}_{\textrm{odd}} \biggl\{ \biggl[ \frac{\cos\theta}{\sqrt{2}} \bigl( \partial_x u_x + \partial_y u_y \bigr) + \sin\theta \,\partial_z u_z \biggr] {\bsf D}_2 - \biggl[ \cos\theta \,\partial_z u_z - \frac{\sin\theta}{\sqrt{2}} \bigl( \partial_x u_x + \partial_y u_y \bigr) \biggr] {\bsf D}_1 \biggr\} \\
    & + B^{2_1}_{0} \biggl\{ \frac{1}{\sqrt{2}} \bigl( \partial_z u_x + \partial_x u_z \bigr) {\bsf S}_3 + \frac{1}{\sqrt{2}} \bigl( \partial_z u_y + \partial_y u_z \bigr) {\bsf S}_4 \biggr\} + B^{2_1}_{\textrm{odd}} \biggl\{ \frac{1}{\sqrt{2}} \bigl( \partial_z u_x + \partial_x u_z \bigr) {\bsf S}_4 - \frac{1}{\sqrt{2}} \bigl( \partial_z u_y + \partial_y u_z \bigr) {\bsf S}_3 \biggr\} \\ 
    & + B^{2_2}_{0} \biggl\{ \frac{1}{\sqrt{2}} \bigl( \partial_x u_x - \partial_y u_y \bigr) {\bsf S}_5 + \frac{1}{\sqrt{2}} \bigl( \partial_x u_y + \partial_y u_x \bigr) {\bsf S}_6 \biggr\} + B^{2_2}_{\textrm{odd}} \biggl\{ \frac{1}{\sqrt{2}} \bigl( \partial_x u_x - \partial_y u_y \bigr) {\bsf S}_6 - \frac{1}{\sqrt{2}} \bigl( \partial_x u_y + \partial_y u_x \bigr) {\bsf S}_5 \biggr\} .
\end{split}
\end{equation}
\end{widetext}
This presentation emphasises the structure in terms of the irreps of the stress and strain under the action of the symmetry group. For apolar materials with $D_{\infty}$ symmetry $B^{2_2}_{\textrm{odd}}=B^{2_1}_{\textrm{odd}}=0$. 
We also record the relation between the moduli here and those in the free energy~\eqref{eq:Fsol} 
\begin{gather}
    B_1 = B^{1\oplus 1}_0 - B^{1\oplus 1}_{1} \cos 2\theta , \\
    B_2 = \frac{1}{\sqrt{2}} \,B^{1\oplus 1}_{1} \sin 2\theta , \\
    B_3 = \frac{1}{2} \Bigl( B^{1\oplus 1}_0 + B^{1\oplus 1}_{1} \cos 2\theta - B^{2_2}_0 \Bigr) , \\
    B_4 = \frac{1}{2} B^{2_2}_0 , \\
    B_5 = B^{2_1}_0 .
\end{gather}

The extra symmetry of the elastomer impacts the elasticity of the $2_1$ irrep. Elastic strains within this irrep that are accompanied by a rotation---independently of the elastic medium---of the polarisation, or uniaxial direction, correspond to symmetries of the elastomer, as shown in Fig.~\ref{fig:softmode}. Since they are symmetries they do not induce any stress and so the moduli $B^{2_1}_0$ and $B^{2_1}_{\rm odd}$ must both vanish in the elastomer.  

In general, a uniaxial solid has three odd elasticities---contributions to the stress-strain relation that violate Maxwell-Betti reciprocity---one associated to each of the distinct representations of the polar uniaxial symmetry group on the spaces of stress and linear strain. Two of these, $B^{2_1}_{\rm odd}$ and $B^{2_2}_{\rm odd}$, are both chiral and polar, i.e. they require $C_{\infty}$ symmetry. The third, $B^{1\oplus 1}_{\rm odd}$, is also allowed in achiral and apolar materials with $D_{\infty h}$ symmetry.  

The odd contribution to the elastic force density is 
\begin{equation}
    \begin{split}
        \boldsymbol{\cal F}_o & = \frac{1}{\sqrt{2}} B^{1\oplus 1}_{\rm odd} \Bigl[ \partial_z \bigl( \partial_x u_x + \partial_y u_y \bigr) \,\hat{\bf z} - \nabla_{\perp} \partial_z u_z \Bigr] \\
        & \quad + \frac{1}{2} B^{2_1}_{\rm odd} \Bigl[ \partial_z \bigl( \partial_y u_x - \partial_x u_y \bigr) \,\hat{\bf z} - \partial_z^2 \,{\bm \varepsilon} \cdot {\bf u}_{\perp} \\
        & \qquad - {\bm \varepsilon} \cdot \nabla_{\perp} \partial_z u_z \Bigr] - \frac{1}{2} B^{2_2}_{\rm odd} \nabla_{\perp}^2 {\bm \varepsilon} \cdot {\bf u}_{\perp} \,.
    \end{split}
\end{equation}
This can be written in Fourier space as a cross product $\boldsymbol{\cal F}_o = {\bm \Omega}_o \times {\bf u}$ with odd rotation vector 
\begin{equation}
    \begin{split}
        {\bm \Omega}_o & = \frac{B^{1\oplus 1}_{\rm odd}}{\sqrt{2}} \bigl( q_x q_z \,\hat{\bf y} - q_y q_z \,\hat{\bf x} \bigr) + \frac{B^{2_1}_{\rm odd}}{2} \Bigl( q_x q_z \,\hat{\bf x} \\
        & \quad + q_y q_z \,\hat{\bf y} - q_z^2 \,\hat{\bf z} \Bigr) - \frac{B^{2_2}_{\rm odd}}{2} \,q_{\perp}^2 \,\hat{\bf z} \,.
    \end{split}
\end{equation}
In the elastomer, where $B^{2_1}_{\rm odd} = 0$ for symmetry reasons, the odd force density vanishes for wavevectors with $q_{\perp} = 0$, whereas in the solid the odd force density is generically non-zero for all wavevectors. This appears in the spectrum in the mode degeneracy of elastomers for wavevectors along the polar axis described in Sec.~\ref{elasin} and~\eqref{eigen_iner}. The degeneracy is lifted by the breaking of elastomeric symmetry in the solid as described in Sec.~\ref{subsec:inertial_solids}. 
Further, as remarked in~\cite{short} $\Omega_{o z}$ has consistent sign if $B^{2_1}_{\rm odd}$ and $B^{2_2}_{\rm odd}$ have the same sign, but is both positive and negative if the two odd moduli are oppositely signed and this represents a topological distinction in the odd-elastic force density~\cite{OddTopology}.

We remark that the same representation-theoretic arguments apply to the description of viscosity (Sec.~\ref{sec:vforceden}), with the linearised strain replaced by the symmetric part of the velocity gradient $\tfrac{1}{2} ( \partial_i v_j + \partial_j v_i )$. If the velocity field is incompressible the isotropic part of the strain (spanned by ${\bsf S}_1$) vanishes and the relation between strain and stress restricts to the deviatoric subspace. In this case~\eqref{eq:elasticity_1+1} should be replaced with a single viscosity relating the ${\bsf S}_2$ subspaces of strain and stress~\cite{SJ_col}.

\section{Chiral active solids in external fields and structured environments}
\label{sec:extenviron}

\numberwithin{equation}{section}
\renewcommand{\theequation}{\thesection.\arabic{equation}}

The uniaxial solids we discussed in the last section were still invariant under rigid rotations. However, we can consider even less symmetric situations. For instance, when a solid moves in an external field along $\hat{{\bf z}}$, the only invariance is under rotations about the $\hat{{\bf z}}$ axis. We can write the equation of motion of a solid moving through a frictional medium in this case as
\begin{widetext}
	\begin{equation}
		\label{Rangan3dchivel}
v_i=a^c_1\epsilon_{ijz}\partial_zu_j+a^c_2\epsilon_{ijz}\partial_j{ u}_z+a_0\delta_{iz}+(a_1\partial_zu_z+a_2\partial_ju_j)\delta_{iz}+a_3\partial_iu_z+a_4\partial_zu_i+\mathcal{O}(\nabla^2)\,.
	\end{equation}
\end{widetext}
The achiral terms $a_i$ are odd in $\hat{{\bf z}}$, while those that require the system to be composed of chiral elements, $a^c_{1,2}$, are even in $\hat{{\bf z}}$. Thus, we see that the breaking of rotation symmetry allows for an additional chiral contribution compared to Sec.~\ref{dynsolfric}. Equation~\eqref{Rangan3dchivel} implies the following equations for displacement fluctuation in a co-moving reference frame $v_0=a_0\hat{z}$:
\begin{equation}
	\partial_t u_x=a_3\partial_x u_z+a_4\partial_z u_x+a_1^c\partial_z u_y+a_2^c\partial_yu_z+\mathcal{O}(\nabla^2)\,,
\end{equation}
\begin{equation}
	\partial_t u_y=a_3\partial_y u_z+a_4\partial_z u_y-a_1^c\partial_z u_x-a_2^c\partial_xu_z+\mathcal{O}(\nabla^2)\,,
\end{equation}
\begin{equation}
	\partial_t u_z=a_2(\partial_xu_x+\partial_yu_y)+(a_1+a_2+a_3+a_4)\partial_z u_z+\mathcal{O}(\nabla^2)\,.
\end{equation}
These equations also imply a generic instability that can again be seen most clearly by looking at perturbations purely along $\hat{{\bf z}}$.

\begin{figure}[t]
    \centering
    \includegraphics[width=0.9\linewidth]{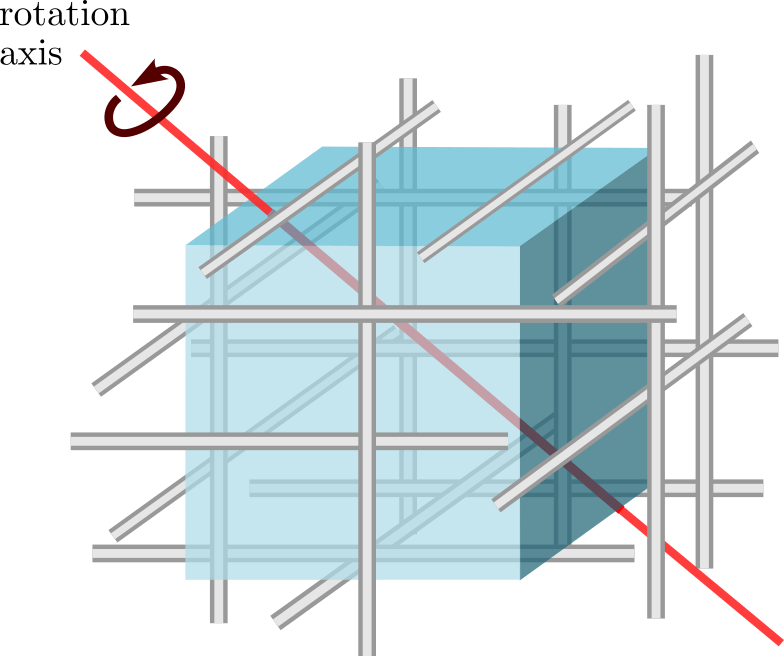}
    \caption{Schematic of an active chiral solid (blue) in an external medium with cubic anisotropy, indicated by the grey mesh. A rotation of the solid about a general axis, shown in red, leads to a destabilising chiral active force density as described in the text.}
    \label{fig:cubic}
\end{figure}

We can also have situations in which there is no axis about which rotation leaves the system invariant. The simplest example of this is a solid in an external medium with cubic anisotropy; see Fig.~\ref{fig:cubic}. When such an active solid is chiral, it generically has a force density $a_3^c\nabla\times{\bf u}$. That is, a rigid rotation of the elastic body with respect to the structured medium leads to an active force density if the material is chiral. If the solid does not have any additional anisotropy, the displacement field dynamics to leading order in gradients is simply
\begin{equation}
    \partial_t{\bf u}=a_3^c\nabla\times{\bf u}+\mathcal{O}\left(\nabla^2\right)\,.
\end{equation}
This implies that there are only two eigenfrequencies at first order in wavenumbers (the third eigenfrequency scales as $\sim q^2$): $\omega_\pm=\pm i a_3^c q$. Note that since the chiral and active force density in this case depends on the curl of the displacement field, the eigenfrequency doesn't depend on the \emph{direction} of the wavevector, i.e., a perturbation along any wavevector direction is unstable.

\section{Conclusion}
\label{sec:conc}
In this article, we have examined the effect of chiral activity on uniaxial elastomeric and solid phases. Uniaxial elastomers are invariant under independent rotation of the alignment direction and elastic material, and we have shown that this symmetry exceptionally stabilises the elastomeric state compared to a uniaxial solid. Further, while the constraints of rotation symmetry restrict the number of chiral odd elasticities to two in three-dimensional polar and chiral active solids, the additional operation that elastomers are invariant under further restricts this number to one in polar and chiral active elastomers. This implies that, as in passive materials, the linearised displacement fluctuations $u_z$ along the polar direction, and ${\bf u}_\perp$ transverse to it, of a chiral active elastomer continue to behave like the displacement fluctuations of a polar and chiral active smectic in three dimensions~\cite{SJ_smec, AM_chi_rev} and a polar and chiral active columnar material~\cite{SJ_col}, respectively, when momentum is conserved. In particular, the chiral odd elasticity that characterises a polar and chiral active elastomeric state is exactly the one that we uncovered earlier in polar and chiral columnar systems~\cite{SJ_col}; at this order in gradients, polar and chiral active smectics do not have an odd elastic force density. In addition to chiral odd elasticities, our formulation of the uniaxial (not necessarily polar) active elastomeric dynamics naturally generates an achiral odd elasticity previously discussed in~\cite{old_odd, AM_sol, shankar2024active}.

Following our discussion of the elastomeric state, we construct the experimentally more relevant active dynamics of uniaxial, chiral \emph{solids} by explicitly imposing a field that breaks the higher elastomeric symmetry but maintains the invariance under rigid rotations. We show that the lower symmetry of this more ordered state allows chiral activity to be much more consequential. Indeed, in a frictional external medium, a uniaxial solid is generically destabilised by chiral activity. The force density responsible for this instability does not exist in an elastomeric state and vanishes linearly with the field breaking the higher symmetry of that phase. Further, as expected, polar active solids have two independent chiral odd elastic moduli instead of one. We show that the presence of the additional chiral force density renders inertial polar active solids generically unstable. This demonstrates that chiral activity is particularly efficient in destroying a polar solid state in multiple dynamical regimes. 

We also discuss solid states in which even the symmetry under rigid rotations is broken due to the presence of external fields or a structured medium. We show that chiral activity tends to destabilise all such phases.

Our work clarifies the effect of chiral activity on a wide range of active solid and elastomeric phases and in a variety of dynamical regimes. While the physics of three-dimensional chiral active solids had not been examined in any detail earlier, this work essentially constitutes a complete discussion of the linear theory of such systems. We show that the chirality allows for richer physics in three-dimensional solid materials than in two dimensions~\cite{AM_hex, scheibner2020odd, fruchart2023odd, AM_chi_rev}. 

As discussed more fully in the accompanying Letter~\cite{short}, our work is of relevance to a variety of systems ranging from tissues to driven skyrmions in liquid crystals.

\begin{acknowledgments} 
	AM acknowledges a TALENT fellowship awarded by CY Cergy Paris Universit\'e and the support of ANR through the grant PSAM. AM, SJK and
	SR thank the Isaac Newton Institute for Mathematical Sciences, Cambridge, for support and hospitality during the programmes ``New Statistical Physics in Living Matter: non-equilibrium states under Adaptive Control'' and ``Anti-diffusive dynamics: from sub-cellular to astrophysical scales'' (EPSRC grant no. EP/R014604/1), and SR additionally for a Rothschild Distinguished Visiting Fellowship. This research was supported in part by grant NSF PHY-2309135 to the Kavli Institute for Theoretical Physics. SR was supported in part by a J C Bose Fellowship of the ANRF, India. SJK was supported by the INI-Simons research fellowship, Cambridge. 
\end{acknowledgments}

\bibliography{ref}

@article{Bouligand_LC,
  title={Liquid crystals and biological morphogenesis: Ancient and new questions},
  author={Bouligand, Yves},
  journal={Comptes Rendus Chimie},
  volume={11},
  number={3},
  pages={281--296},
  year={2008}
}

@article{Bouligand_twisted,
  title={Twisted fibrous arrangements in biological materials and cholesteric mesophases},
  author={Bouligand, Y},
  journal={Tissue and Cell},
  volume={4},
  number={2},
  pages={189--217},
  year={1972},
  publisher={Elsevier}
}

@inproceedings{Heartstrings,
  title={The structure and function of the helical heart and its buttress wrapping. I. The normal macroscopic structure of the heart},
  author={Torrent-Guasp, Francisco and Buckberg, Gerald D and Clemente, Carmine and Cox, James L and Coghlan, H Cecil and Gharib, Morteza},
  booktitle={Seminars in Thoracic and Cardiovascular Surgery},
  volume={13},
  number={4},
  pages={301},
  year={2001},
  organization={Elsevier}
}

@article{Mermin_blue,
  title={Crystalline liquids: the blue phases},
  author={Wright, David C and Mermin, N David},
  journal={Reviews of Modern Physics},
  volume={61},
  number={2},
  pages={385},
  year={1989},
  publisher={APS}
}

@article{Das_Barma,
  title={Particles sliding on a fluctuating surface: phase separation and power laws},
  author={Das, Dibyendu and Barma, Mustansir},
  journal={Physical Review Letters},
  volume={85},
  number={8},
  pages={1602},
  year={2000},
  publisher={APS}
}

@article{AM_interface,
  title={Two-dimensional long-range uniaxial order in three-dimensional active fluids},
  author={Maitra, Ananyo},
  journal={Nature Physics},
  volume={19},
  number={5},
  pages={733--740},
  year={2023},
  publisher={Nature Publishing Group UK London}
}

@misc{OddTopology,
  commented_title={Topology of Odd Elasticity},
  title={In preparation},
  author={Kole, Sxx J and Alexander, Gareth P and Maitra, Ananyo and Ramaswamy, Sriram},
  year={2026}
}

@article{Echebarria2,
  title={Stability of oscillating hexagons in rotating convection},
  author={Echebarria, Blas and Riecke, Hermann},
  journal={Physica D: Nonlinear Phenomena},
  volume={143},
  number={1-4},
  pages={187--204},
  year={2000},
  publisher={Elsevier}
}

@article{Echebarria1,
  title={Instabilities of hexagonal patterns with broken chiral symmetry},
  author={Echebarria, Blas and Riecke, Hermann},
  journal={Physica D: Nonlinear Phenomena},
  volume={139},
  number={1-2},
  pages={97--108},
  year={2000},
  publisher={Elsevier}
}

@article{Madruga,
  title={Hexagonal patterns in a model for rotating convection},
  author={Madruga, Santiago and P{\'e}rez-Garc{\'\i}a, Carlos},
  journal={International Journal of Bifurcation and Chaos},
  volume={14},
  number={01},
  pages={107--117},
  year={2004},
  publisher={World Scientific}
}

@article{Rayan,
  title={Inertia drives a flocking phase transition in viscous active fluids},
  author={Chatterjee, Rayan and Rana, Navdeep and Simha, R Aditi and Perlekar, Prasad and Ramaswamy, Sriram},
  journal={Physical Review X},
  volume={11},
  number={3},
  pages={031063},
  year={2021},
  publisher={APS}
}

@article{Aditi,
  title={Hydrodynamic fluctuations and instabilities in ordered suspensions of self-propelled particles},
  author={Aditi Simha, R and Ramaswamy, Sriram},
  journal={Physical {R}eview {L}etters},
  volume={89},
  number={5},
  pages={058101},
  year={2002},
  publisher={APS}
}

@article{Deboo_viscel,
  title={Active viscoelasticity of odd materials},
  author={Banerjee, Debarghya and Vitelli, Vincenzo and J{\"u}licher, Frank and Sur{\'o}wka, Piotr},
  journal={Physical Review Letters},
  volume={126},
  number={13},
  pages={138001},
  year={2021},
  publisher={APS}
}

@article{old_odd,
  title={Non-axial muscle stress and stiffness},
  author={Zahalak, George I},
  journal={Journal of {T}heoretical {B}iology},
  volume={182},
  number={1},
  pages={59--84},
  year={1996},
  publisher={Elsevier}
}

@article{khain2022stokes,
  title={Stokes flows in three-dimensional fluids with odd and parity-violating viscosities},
  author={Khain, Tali and Scheibner, Colin and Fruchart, Michel and Vitelli, Vincenzo},
  journal={Journal of Fluid Mechanics},
  volume={934},
  pages={A23},
  year={2022},
  publisher={Cambridge University Press}
}

@article{kole2025non,
  title={Non-reciprocal interactions drive emergent chiral crystallites},
  author={Kole, Sxx J and Chao, Xichen and Mauleon-Amieva, Abraham and Hanai, Ryo and Royall, C Patrick and Liverpool, Tanniemola B},
  journal={arXiv preprint arXiv:2501.15996},
  year={2025}
}

@article{maxwell1864,
  title={On the calculation of the equilibrium and stiffness of frames},
  author={Maxwell, J Clerk},
  journal={The London, Edinburgh, and Dublin Philosophical Magazine and Journal of Science},
  volume={27},
  number={182},
  pages={294--299},
  year={1864},
  publisher={Taylor \& Francis}
}

@article{betti1872,
  title={Teoria della elasticit{\`a}}, 
  author={Betti, Enrico},
  journal={Nuovo Cimento},
  volume={7},
  number={1},
  pages={5--21},
  year={1872},
  publisher={Societ{\`a} Italiana di Fisica Bologna}
}

@article{pisljar2022,
  title={Blue phase {III}: Topological fluid of skyrmions},
  author={Pi{\v{s}}ljar, J and Ghosh, S and Turlapati, S and Rao, NVS and {\v{S}}karabot, M and Mertelj, A and Petelin, A and Nych, A and Marin{\v{c}}i{\v{c}}, M and Pusovnik, A and Ravnik, M and Mu{\v{s}}evi{\v{c}}, I},
  journal={Physical Review X},
  volume={12},
  number={1},
  pages={011003},
  year={2022},
  publisher={APS}
}

@article{renn1988,
  title={Abrikosov dislocation lattice in a model of the cholesteric--to--smectic-{A} transition},
  author={Renn, Scott R and Lubensky, Tom C},
  journal={Physical Review A},
  volume={38},
  number={4},
  pages={2132},
  year={1988},
  publisher={APS}
}

@article{hough2009chiral,
  title={Chiral isotropic liquids from achiral molecules},
  author={Hough, LE and Spannuth, M and Nakata, Michi and Coleman, DA and Jones, CD and Dantlgraber, G and Tschierske, Carsten and Watanabe, J and K{\"o}rblova, E and Walba, DM and others},
  journal={Science},
  volume={325},
  number={5939},
  pages={452--456},
  year={2009},
  publisher={American Association for the Advancement of Science}
}

@article{hough2009helical,
  title={Helical nanofilament phases},
  author={Hough, LE and Jung, Hee-Tae and Kr{\"u}erke, Daniel and Heberling, MS and Nakata, Michi and Jones, CD and Chen, D and Link, Darren R and Zasadzinski, J and Heppke, Gerd and others},
  journal={Science},
  volume={325},
  number={5939},
  pages={456--460},
  year={2009},
  publisher={American Association for the Advancement of Science}
}

@article{matsumoto2017straight,
  title={Straight round the twist: frustration and chirality in smectics-{A}},
  author={Matsumoto, Elisabetta A and Kamien, Randall D and Alexander, Gareth P},
  journal={Interface Focus},
  volume={7},
  number={4},
  pages={20160118},
  year={2017}
}

@article{chen2013,
  title={Generating the {H}opf fibration experimentally in nematic liquid crystals},
  author={Chen, Bryan Gin-ge and Ackerman, Paul J and Alexander, Gareth P and Kamien, Randall D and Smalyukh, Ivan I},
  journal={Physical Review Letters},
  volume={110},
  number={23},
  pages={237801},
  year={2013},
  publisher={APS}
}

@article{machon2017,
  title={Contact topology and the structure and dynamics of cholesterics},
  author={Machon, Thomas},
  journal={New Journal of Physics},
  volume={19},
  number={11},
  pages={113030},
  year={2017},
  publisher={IOP Publishing}
}

@article{pollard2019,
  title={Point defects, topological chirality, and singularity theory in cholesteric liquid-crystal droplets},
  author={Pollard, Joseph and Posnjak, Gregor and {\v{C}}opar, Simon and Mu{\v{s}}evi{\v{c}}, Igor and Alexander, Gareth P},
  journal={Physical Review X},
  volume={9},
  number={2},
  pages={021004},
  year={2019},
  publisher={APS}
}

@article{pollard2023,
  title={Contact topology and the classification of disclination lines in cholesteric liquid crystals},
  author={Pollard, Joseph and Alexander, Gareth P},
  journal={Physical Review Letters},
  volume={130},
  number={22},
  pages={228102},
  year={2023},
  publisher={APS}
}

@article{tai2019,
  title={Three-dimensional crystals of adaptive knots},
  author={Tai, Jung-Shen B and Smalyukh, Ivan I},
  journal={Science},
  volume={365},
  number={6460},
  pages={1449--1453},
  year={2019},
  publisher={American Association for the Advancement of Science}
}

@article{wu2022hopfions,
  title={Hopfions, heliknotons, skyrmions, torons and both abelian and nonabelian vortices in chiral liquid crystals},
  author={Wu, Jin-Sheng and Smalyukh, Ivan I},
  journal={Liquid Crystals Reviews},
  volume={10},
  number={1-2},
  pages={34--68},
  year={2022},
  publisher={Taylor \& Francis}
}

@article{roessler2006spontaneous,
  title={Spontaneous skyrmion ground states in magnetic metals},
  author={R{\"o}{\ss}ler, Ulrich K and Bogdanov, AN and Pfleiderer, C},
  journal={Nature},
  volume={442},
  number={7104},
  pages={797--801},
  year={2006},
  publisher={Nature Publishing Group UK London}
}

@article{sutcliffe2018hopfions,
  title={Hopfions in chiral magnets},
  author={Sutcliffe, Paul},
  journal={Journal of Physics A: Mathematical and Theoretical},
  volume={51},
  number={37},
  pages={375401},
  year={2018},
  publisher={IOP Publishing}
}

@article{kent2021creation,
  title={Creation and observation of Hopfions in magnetic multilayer systems},
  author={Kent, Noah and Reynolds, Neal and Raftrey, David and Campbell, Ian TG and Virasawmy, Selven and Dhuey, Scott and Chopdekar, Rajesh V and Hierro-Rodriguez, Aurelio and Sorrentino, Andrea and Pereiro, Eva and others},
  journal={Nature Communications},
  volume={12},
  number={1},
  pages={1562},
  year={2021},
  publisher={Nature Publishing Group UK London}
}

@article{chen2026laser,
  title={Laser-induced nucleation of magnetic hopfions},
  author={Chen, Xiaowen and Yang, Donghai and Li, Zefang and Guo, Jiangteng and Wang, Haixue and Hu, Yue and Kuchkin, Vladyslav M and Savchenko, Andrii S and Zhang, Huai and Ding, Bei and others},
  journal={Nature Physics},
  volume={22},
  pages={736--744},
  year={2026},
  publisher={Nature Publishing Group UK London}
}

@article{li2026electrically,
  title={Electrically writing a magnetic heliknoton in a chiral magnet},
  author={Li, Long and Song, Dongsheng and Wang, Weiwei and Kong, Lingyao and Zhang, Shuisen and Wang, Ning and Zhang, Shilei and Tian, Mingliang and Zang, Jiadong and Liu, Yizhou and others},
  journal={Nature Materials},
  volume={25},
  pages={950--955},
  year={2026},
  publisher={Nature Publishing Group UK London}
}

@article{bililign2022motile,
  title={Motile dislocations knead odd crystals into whorls},
  author={Bililign, Ephraim S and Balboa Usabiaga, Florencio and Ganan, Yehuda A and Poncet, Alexis and Soni, Vishal and Magkiriadou, Sofia and Shelley, Michael J and Bartolo, Denis and Irvine, William Txx M},
  journal={Nature Physics},
  volume={18},
  number={2},
  pages={212--218},
  year={2022},
  publisher={Nature Publishing Group UK London}
}

@article{veenstra2025adaptive,
  title={Adaptive locomotion of active solids},
  author={Veenstra, Jonas and Scheibner, Colin and Brandenbourger, Martin and Binysh, Jack and Souslov, Anton and Vitelli, Vincenzo and Coulais, Corentin},
  journal={Nature},
  volume={639},
  number={8056},
  pages={935--941},
  year={2025},
  publisher={Nature Publishing Group UK London}
}

@article{liebchen2022chiral,
  title={Chiral active matter},
  author={Liebchen, Benno and Levis, Demian},
  journal={Europhysics Letters},
  volume={139},
  number={6},
  pages={67001},
  year={2022},
  publisher={EDP Sciences, IOP Publishing and Societ{\`a} Italiana di Fisica}
}

@article{AM_sol,
  title={Oriented active solids},
  author={Maitra, Ananyo and Ramaswamy, Sriram},
  journal={Physical {R}eview {L}etters},
  volume={123},
  number={23},
  pages={238001},
  year={2019},
  publisher={APS}
}

@book{Kelvin,
  title={Baltimore lectures on molecular dynamics and the wave theory of light},
  author={Kelvin, Lord William Thomson},
  year={1904},
  publisher={CUP Archive}
}

@article{scheibner2020odd,
  title={Odd elasticity},
  author={Scheibner, Colin and Souslov, Anton and Banerjee, Debarghya and Sur{\'o}wka, Piotr and Irvine, William Txx M and Vitelli, Vincenzo},
  journal={Nature Physics},
  volume={16},
  number={4},
  pages={475--480},
  year={2020},
  publisher={Nature Publishing Group UK London}
}

@article{maitra2019spontaneous,
  title={Spontaneous rotation can stabilise ordered chiral active fluids},
  author={Maitra, Ananyo and Lenz, Martin},
  journal={Nature {C}ommunications},
  volume={10},
  number={1},
  pages={920},
  year={2019},
  publisher={Nature Publishing Group UK London}
}

@article{fruchart2023odd,
  title={Odd viscosity and odd elasticity},
  author={Fruchart, Michel and Scheibner, Colin and Vitelli, Vincenzo},
  journal={Annual Review of Condensed Matter Physics},
  volume={14},
  number={1},
  pages={471--510},
  year={2023},
  publisher={Annual Reviews}
}

@article{AM_Ladoux_PNAS,
  title={Flocking and giant fluctuations in epithelial active solids},
  author={Shen, Yuan and O’Byrne, J{\'e}r{\'e}my and Schoenit, Andreas and Maitra, Ananyo and M{\`e}ge, Ren{\'e}-Marc and Voituriez, Rapha{\"e}l and Ladoux, Benoit},
  journal={Proceedings of the National Academy of Sciences},
  volume={122},
  number={16},
  pages={e2421327122},
  year={2025},
  publisher={National Academy of Sciences}
}

@article{Fernando_mem,
  title={Interface dynamics of wet active systems},
  author={Caballero, Fernando and Maitra, Ananyo and Nardini, Cesare},
  journal={Physical Review Letters},
  volume={134},
  number={8},
  pages={087105},
  year={2025},
  publisher={APS}
}

@article{Stenull_lub_hydro,
  title={Dynamics of nematic elastomers},
  author={Stenull, Olaf and Lubensky, Tom C},
  journal={Physical Review E—Statistical, Nonlinear, and Soft Matter Physics},
  volume={69},
  number={5},
  pages={051801},
  year={2004},
  publisher={APS}
}

@book{deGen,
  title={The Physics of Liquid Crystals},
  author={De Gennes, Pierre-Gilles and Prost, Jacques},
  year={1993},
  publisher={Oxford University Press, Oxford}
}

@book{LandauLifshitz,
  title={Theory of Elasticity, 3rd Edition},
  author={Landau, L.~D. and Lifshitz, E.~M.},
  year={1986},
  publisher={Butterworth-Heinemann, Oxford}
}

@article{AM_floc,
  title={The inconvenient truth about flocks},
  author={Chen, Leiming and Jentsch, Patrick and Lee, Chiu Fan and Maitra, Ananyo and Ramaswamy, Sriram and Toner, John},
  journal={arXiv preprint arXiv:2503.17064},
  year={2025}
}

@article{ramaswamy2010mechanics,
  title={The mechanics and statistics of active matter},
  author={Ramaswamy, Sriram},
  journal={Annu. Rev. Condens. Matter Phys.},
  volume={1},
  number={1},
  pages={323--345},
  year={2010},
  publisher={Annual Reviews}
}

@article{lubensky1985hydrodynamics,
  title={Hydrodynamics of icosahedral quasicrystals},
  author={Lubensky, Tom C and Ramaswamy, Sriram and Toner, John},
  journal={Physical Review B},
  volume={32},
  number={11},
  pages={7444},
  year={1985},
  publisher={APS}, 
  url={https://journals.aps.org/prb/abstract/10.1103/PhysRevB.32.7444}
}

@article{zahalak1996non,
  title={Non-axial muscle stress and stiffness},
  author={Zahalak, George I},
  journal={Journal of Theoretical Biology},
  volume={182},
  number={1},
  pages={59--84},
  year={1996},
  publisher={Elsevier}, 
  url={https://doi.org/10.1006/jtbi.1996.0143}
}

@article{AM_sol_err,
  title = {Erratum: Oriented Active Solids [Phys. Rev. Lett. 123, 238001 (2019)]},
  author = {Maitra, Ananyo and Ramaswamy, Sriram},
  journal = {Phys. Rev. Lett.},
  volume = {130},
  issue = {26},
  pages = {269901},
  numpages = {1},
  year = {2023},
  month = {Jun},
  publisher = {American Physical Society},
  doi = {10.1103/PhysRevLett.130.269901},
  url = {https://link.aps.org/doi/10.1103/PhysRevLett.130.269901}
}

@article{RJoy,
  title={Nonreciprocal constitutive laws for oriented active solids},
  author={N{\'e}meth, Bal{\'a}zs and Kobayashi, Takuya and Adhikari, Ronojoy},
  journal={arXiv preprint arXiv:2509.11430},
  year={2025}
}

@book{Chaikin,
  title={Principles of Condensed Matter Physics},
  author={Chaikin, Paul M and Lubensky, Tom C},
  year={1995},
  publisher={Cambridge University Press, Cambridge}
}

@article{SJ_smec,
  title={Layered chiral active matter: Beyond odd elasticity},
  author={Kole, Sxx J and Alexander, Gareth P and Ramaswamy, Sriram and Maitra, Ananyo},
  journal={Physical Review Letters},
  volume={126},
  number={24},
  pages={248001},
  year={2021},
  publisher={APS}
}

@article{SJ_col,
  title={Chirality and odd mechanics in active columnar phases},
  author={Kole, Sxx J and Alexander, Gareth P and Maitra, Ananyo and Ramaswamy, Sriram},
  journal={PNAS {N}exus},
  volume={3},
  number={10},
  pages={pgae398},
  year={2024},
  publisher={Oxford University Press US}
}

@article{Mermin,
  title={Crystalline order in two dimensions},
  author={Mermin, N David},
  journal={Physical Review},
  volume={176},
  number={1},
  pages={250},
  year={1968},
  publisher={APS}
}

@article{rustarazo2026adhesion,
  title={Adhesion-driven rigidity transition decoupled from density-driven jamming triggers epithelial organization in embryonic tissues},
  author={Rustarazo-Calvo, Laura and Pallares-Cartes, Cristina and Aguirre-Tamaral, Adri{\'a}n and Floris, Elisa and Hingerl, Maximilian and Autorino, Camilla and Khan, Arif Ul Maula and Corominas-Murtra, Bernat and Petridou, Nicoletta I},
  journal={Nature Physics},
  volume={22},
  pages={941--953},
  year={2026},
  publisher={Nature Publishing Group UK London}
}

@article{spatarelu2019biomechanics,
  title={Biomechanics of collective cell migration in cancer progression: experimental and computational methods},
  author={Spatarelu, Catalina-Paula and Zhang, Hao and Nguyen, Dung Trung and Han, Xinyue and Liu, Ruchuan and Guo, Qiaohang and Notbohm, Jacob and Fan, Jing and Liu, Liyu and Chen, Zi},
  journal={ACS {B}iomaterials {S}cience \& {E}ngineering},
  volume={5},
  number={8},
  pages={3766},
  year={2019}
}

@article{rorth2009collective,
  title={Collective cell migration},
  author={R{\o}rth, Pernille},
  journal={Annual {R}eview of {C}ell and {D}evelopmental},
  volume={25},
  pages={407--429},
  year={2009},
  publisher={Annual Reviews}
}

@article{ellison2016cell,
  title={Cell--cell communication enhances the capacity of cell ensembles to sense shallow gradients during morphogenesis},
  author={Ellison, David and Mugler, Andrew and Brennan, Matthew D and Lee, Sung Hoon and Huebner, Robert J and Shamir, Eliah R and Woo, Laura A and Kim, Joseph and Amar, Patrick and Nemenman, Ilya and others},
  journal={Proceedings of the National Academy of Sciences},
  volume={113},
  number={6},
  pages={E679--E688},
  year={2016},
  publisher={National Academy of Sciences}
}

@article{mugler2016limits,
  title={Limits to the precision of gradient sensing with spatial communication and temporal integration},
  author={Mugler, Andrew and Levchenko, Andre and Nemenman, Ilya},
  journal={Proceedings of the National Academy of Sciences},
  volume={113},
  number={6},
  pages={E689--E695},
  year={2016},
  publisher={National Academy of Sciences}
}

@article{camley2016emergent,
  title={Emergent collective chemotaxis without single-cell gradient sensing},
  author={Camley, Brian A and Zimmermann, Juliane and Levine, Herbert and Rappel, Wouter-Jan},
  journal={Physical {R}eview {L}etters},
  volume={116},
  number={9},
  pages={098101},
  year={2016},
  publisher={APS}
}

@article{Norio,
  title={Buckling, stiffening, and negative dissipation in the dynamics of a biopolymer in an active medium},
  author={Kikuchi, Norio and Ehrlicher, Allen and Koch, Daniel and K{\"a}s, Josef A and Ramaswamy, Sriram and Rao, Madan},
  journal={Proceedings of the National Academy of Sciences},
  volume={106},
  number={47},
  pages={19776--19779},
  year={2009},
  publisher={National Academy of Sciences}
}

@article{Toner_rot,
  title={Only the ambidextrous can flock: Two-dimensional chiral {M}althusian flocks, time cholesterics, and the {K}ardar-{P}arisi-{Z}hang equation},
  author={Chen, Leiming and Lee, Chiu Fan and Toner, John},
  journal={Physical Review E},
  volume={113},
  number={4},
  pages={045419},
  year={2026},
  publisher={APS}
}

@article{Lubensy_chol,
  title={Hydrodynamics of cholesteric liquid crystals},
  author={Lubensky, Tom C},
  journal={Physical Review A},
  volume={6},
  number={1},
  pages={452},
  year={1972},
  publisher={APS}
}

@article{Alexander_chol,
  title = {Screw symmetry, chiral hydrodynamics, and odd instability in active cholesterics},
  author = {Alexander, Gareth P. and Kole, Sxx J and Maitra, Ananyo and Ramaswamy, Sriram},
  journal = {Phys. Rev. E},
  volume = {112},
  issue = {5},
  pages = {055424},
  numpages = {11},
  year = {2025},
  month = {Nov},
  publisher = {American Physical Society},
}

@article{Oseen_LC,
author ="Oseen, C. W.",
title  ="The theory of liquid crystals",
journal  ="Trans. Faraday Soc. ",
year  ="1933",
volume  ="29",
issue  ="140",
pages  ="883-899",
publisher  ="The Royal Society of Chemistry"}

@article{Frank_LC,
author ="Frank, F. C.",
title  ="I. {L}iquid crystals. {O}n the theory of liquid crystals",
journal  ="Discuss. Faraday Soc.",
year  ="1958",
volume  ="25",
issue  ="0",
pages  ="19-28",
publisher  ="The Royal Society of Chemistry"}

@article{shankar2024active,
  title={Active hydraulics and odd elasticity of muscle fibres},
  author={Shankar, Suraj and Mahadevan, Lakshminarayanan},
  journal={Nature Physics},
  volume={20},
  number={9},
  pages={1501--1508},
  year={2024},
  publisher={Nature Publishing Group UK London}
}

@article{RMP,
  title={Hydrodynamics of soft active matter},
  author={Marchetti, M Cristina and Joanny, Jean-Fran{\c{c}}ois and Ramaswamy, Sriram and Liverpool, Tanniemola B and Prost, Jacques and Rao, Madan and Simha, R Aditi},
  journal={Reviews of {M}odern {P}hysics},
  volume={85},
  number={3},
  pages={1143--1189},
  year={2013},
  publisher={APS}
}

@article{AM_hex,
  title={Chiral active hexatics: Giant number fluctuations, waves, and destruction of order},
  author={Maitra, Ananyo and Lenz, Martin and Voituriez, Raphael},
  journal={Physical Review Letters},
  volume={125},
  number={23},
  pages={238005},
  year={2020},
  publisher={APS}
}

@article{AM_chi_rev,
  title={Activity Unmasks Chirality in Liquid-Crystalline Matter},
  author={Maitra, Ananyo},
  journal={Annual Review of Condensed Matter Physics},
  volume={16},
  pages={275--295},
  year={2025},
  publisher={Annual Reviews}
}

@article{Olmsted,
  title={Rotational invariance and Goldstone modes in nematic elastomers and gels},
  author={Olmsted, Peter D},
  journal={Journal de Physique II},
  volume={4},
  number={12},
  pages={2215--2230},
  year={1994},
  publisher={EDP Sciences}
}

@article{Ranjan_Lubensky,
  title={Symmetries and elasticity of nematic gels},
  author={Lubensky, Tom C and Mukhopadhyay, Ranjan and Radzihovsky, Leo and Xing, Xiangjun},
  journal={Physical {R}eview {E}},
  volume={66},
  number={1},
  pages={011702},
  year={2002},
  publisher={APS}
}

@book{Warner_Terentjev,
  title={Liquid crystal elastomers},
  author={Warner, Mark and Terentjev, Eugene Michael},
  volume={120},
  year={2007},
  publisher={Oxford {U}niversity {P}ress, {O}xford}
}

@article{Xing_annals,
  title={Nonlinear elasticity, fluctuations and heterogeneity of nematic elastomers},
  author={Xing, Xiangjun and Radzihovsky, Leo},
  journal={Annals of Physics},
  volume={323},
  number={1},
  pages={105--203},
  year={2008},
  publisher={Elsevier}
}

@article{Rangan,
  title={Are steadily moving crystals unstable?},
  author={Lahiri, Rangan and Ramaswamy, Sriram},
  journal={Physical {R}eview {L}etters},
  volume={79},
  number={6},
  pages={1150},
  year={1997},
  publisher={APS}
}

@misc{AMCMSR,
  title={To be published},
  author={Maitra, Ananyo and Marchetti, Cristina and Ramaswamy, Sriram},
}

@misc{short,
  title={Accompanying {L}etter},
  author={Kole, Sxx J and Alexander, Gareth P and Maitra, Ananyo and Ramaswamy, Sriram},
  year={2026}
}

@article{Ano_apol,
  title={A nonequilibrium force can stabilize 2D active nematics},
  author={Maitra, Ananyo and Srivastava, Pragya and Marchetti, M Cristina and Lintuvuori, Juho S and Ramaswamy, Sriram and Lenz, Martin},
  journal={Proceedings of the National Academy of Sciences},
  volume={115},
  number={27},
  pages={6934--6939},
  year={2018},
  publisher={National Acad Sciences}
}

@article{Ano_pol,
  title={Swimmer suspensions on substrates: anomalous stability and long-range order},
  author={Maitra, Ananyo and Srivastava, Pragya and Marchetti, M Cristina and Ramaswamy, Sriram and Lenz, Martin},
  journal={Physical Review Letters},
  volume={124},
  number={2},
  pages={028002},
  year={2020},
  publisher={APS}
}

\end{document}